\documentclass[twocolumn,showpacs,superscriptaddress,amsmath,amssymb]{revtex4}
\usepackage{graphicx}% Include figure files
\usepackage{dcolumn}% Align table columns on decimal point
\usepackage{bm}% bold math
\usepackage{color}
\usepackage{mathrsfs}
\usepackage{ulem}
\usepackage[german]{babel}

\def\bd{
\begin{document}} \def\ed{\end{document}}
\def\bmp{\begin{minipage}} \def\emp{\end{minipage}}
\def\bcc{\begin{center}} \def\ecc{\end{center}}     \def\npg{\newpage}
\def\beq{\begin{equation}} \def\eeq{\end{equation}} \def\hph{\hphantom}
\def\be{\begin{equation}} \def\ee{\end{equation}} \def\r#1{$^{[#1]}$}
\def\n{\noindent} \def\ni{\noindent} \def\pa{\parindent}
\def\hs{\hskip} \def\vs{\vskip} \def\hf{\hfill} \def\ej{\vfill\eject}
\def\cl{\centerline} \def\ob{\obeylines}  \def\ls{\leftskip}
\def\underbar#1{$\setbox0=\hbox{#1} \dp0=1.5pt \mathsurround=0pt
   \underline{\box0}$}   \def\ub{\underbar}    \def\ul{\underline}
\def\f{\left} \def\g{\right} \def\e{{\rm e}} \def\o{\over} \def\d{{\rm d}}
\def\vf{\varphi} \def\pl{\partial} \def\cov{{\rm cov}} \def\ch{{\rm ch}}
\def\la{\langle} \def\ra{\rangle} \def\EE{e$^+$e$^-$} \def\pt{p_{\rm t}}
\def\pti{p_{{\rm t},i}} \def\vti{v_{{\rm t},i}}
\def\ptj{p_{{\rm t},j}}\def\Pt{P_{\rm t}} \def\vt{v_{\rm t}}

\def\bitz{\begin{itemize}} \def\eitz{\end{itemize}}
\def\btbl{\begin{tabular}} \def\etbl{\end{tabular}}
\def\btbb{\begin{tabbing}} \def\etbb{\end{tabbing}}
\def\beqar{\begin{eqnarray}} \def\eeqar{\end{eqnarray}}
\def\\{\hfill\break} \def\dit{\item{-}} \def\i{\item}
\def\bbb{\begin{thebibliography}{9} \itemsep=-1mm}
\def\ebb{\end{thebibliography}} \def\bb{\bibitem}
\def\bpic{\begin{picture}(260,240)} \def\epic{\end{picture}}
\def\akgt{\cl{\bf ACKNOWLEDGMENTS}}
\def\fgn{\noindent{\bf\large\bf figure captions}}
%%%%%%%%%%%%%%%%%%%%%%%%%%%%%%%%%%%%%%%%%%%%%%%%%%%%%%%%%%%%%%%%%%%%%%
\def\m1{\langle N_p\rangle} \def\u2{\langle N_{\bar p}\rangle} \def\Nap{N_{\bar
p}}
%%%%%%%%%%%%%%%%%%%%%%%%%%%%%%%%%%%%%%%%%%%%%%%%%%%%%%%%%%%%%%%%%%%%%%%%%%%%%%
\def\lan{\langle}
\def\ran{\rangle}
\def\p{\pi}
\def\ifmath#1{\relax\ifmmode #1\else $#1$\fi}%
\def\rc{\ifmath{{\mathrm{c}}}}
\def\cut{\ifmath{{\mathrm{cut}}}}
\def\rF{\ifmath{{\mathrm{F}}}}
\def\rK{\ifmath{{\mathrm{K}}}}
\def\rp{\ifmath{{\mathrm{p}}}}
\def\rt{\ifmath{{\mathrm{t}}}}
\def\LAB{\ifmath{{\mathrm{LAB}}}}
\def\cut{\ifmath{{\mathrm{cut}}}}
\def\beq{\begin{equation}}
\def\eeq{\end{equation}}

\newcommand{\cinst}[2]{$^{\mathrm{#1}}$~#2\par}
\newcommand{\crefi}[1]{$^{\mathrm{#1}}$}
\newcommand{\crefii}[2]{$^{\mathrm{#1,#2}}$}
\newcommand{\crefiii}[3]{$^{\mathrm{#1,#2,#3}}$}
\newcommand{\HRule}{\rule{0.5\linewidth}{0.5mm}}

\bd
\title{Fixed point behavior of cumulants in the three-dimensional Ising universality class}

\author{Xue Pan}\email{panxue1624@163.com}
\affiliation{School of Electronic Engineering, Chengdu Technological University, Chengdu 611730, China}

\begin{abstract}
%\sout{}
High-order cumulants and factorial cumulants of conserved charges are suggested to study the critical dynamics in heavy-ion collision experiments. In this paper, using the parametric representation of the three-dimensional Ising model which is believed to belong to the same universality class with the Quantum chromo-dynamics, temperature dependence of the second- to fourth-order (factorial) cumulants of the order parameter is studied. It is found that the values of the normalized cumulants are independent of the external magnetic field at the critical temperature, which results in a fixed point in the temperature dependence of the normalized cumulants. In finite-size systems simulated by Monte Carlo method, the fixed point behavior still exists at the temperature near the critical one. The fixed point behavior is also appeared in the temperature dependence of normalized factorial cumulants at least from the fourth-order one. With a mapping from the Ising model to QCD, the fixed point behavior is also found in the energy dependence of the normalized cumulants (or fourth-order factorial cumulants) along different freeze-out curves.
\end{abstract}


\maketitle

\section{Introduction}
One of the main goals of current relativistic heavy-ion collision experiments is to make clear the phase diagram of quantum chromo-dynamics (QCD)~\cite{main goal}. At vanishing baryon chemical potential, the transition from hadronic matter to quark-gluon plasma has been proved to be a crossover by lattice QCD~\cite{fodor-nature}. Due to the fermion sign problem, lattice QCD can not calculate the cases at large baryon chemical potential. Some effective theories predict that the QCD system undergoes a first order phase transition at high baryon density and low temperature~\cite{first-PRD, first-NPB,first-PRL,first-NPB1,first-LQCD,first-PRC}. From first order phase transition to crossover, there is a critical point, which is a unique feature of the QCD phase diagram. Large fluctuations and correlations of conserved charges are expected at the critical point.

The high-order cumulants of conserved charges, reflecting their fluctuations, are suggested to search for the critical point~\cite{stephanov-prl91, koch, Stephanov-prl102, Karsch-EPJC71}. Results from effective theories of QCD suggest that the non-monotonic behavior of the high-order cumulants are related to the critical point~\cite{Asakawa-prl103, Fuweijie, Vladi}. Especially, sign change of the fourth-order cumulant of net-proton is used to search for the critical point in experiments~\cite{Stephanov-prl107, Phys. Rev. Lett. 112.032302}. While in Ref.~\cite{Chin. Phys. C.43.033103, Chin. Phys. C.45.104103}, the authors argued that the sign change is not sufficient to prove the presence of the critical point. Other work pointed out that the peak structure remains a solid feature and can be used as a clean signature of the critical point~\cite{Eur. Phys. J. C.79.245, PhysRevC103034901}.

Recently, the factorial cumulants, which are also known as the integrated multi-particle correlations, get a lot of attention~\cite{Phys. Lett. B.728.386-392, Nuclear Physics A.942.65-96, Phys. Rev. C.93.034915, Phys. Rev. C.95.064912, Phys. Rev. C.95.054906, Phys. Lett. B.774.623-629, Phys. Rev. C.96.024910, Eur. Phys. J. C.77.288, Phys. Rev. C.100.051902}. Multi-proton correlations have been found in the STAR data, at least at the lower energies~\cite{Phys. Rev. C.95.054906, Phys. Rev. C.98.054901, Eur. Phys. J. C.77.288}. It has been shown that the signs of the second- to fourth-order factorial cumulants are a useful tool to exclude regions in the QCD phase diagram close to the critical point using parametric representation of the Ising model~\cite{Phys. Rev. C.95.054906}. The causes of sign change of factorial cumulants far away from the critical point compared with the cumulants have been analyzed in our recent work~\cite{panx}. In the vicinity of the critical point, the sign and temperature dependence of factorial cumulants is almost the same with that of the cumulants. It has also argued in Ref.~\cite{Phys. Rev. C.93.034915} that the cumulants and factorial cumulants can not be distinguished in the vicinity of the critical point in a model of critical fluctuations.

Except non-monotonic behavior or sign change of the cumulants and factorial cumulants, the other behavior of the high-order cumulant is also suggested searching for the critical point, such as the finite-size scaling~\cite{Phys. Rev. D.97.034015, Journal of Physics G: Nuclear and Particle Physics.42.015104}. The finite-size scaling implies a fixed point. Usually, the fixed point is obtained from the scale transformation of the re-normalization group, resulting in the independence of rescaled thermodynamics on the system sizes at the critical point~\cite{fixed point 1, fixed point 2, fixed point 3}. This feature has also been used to search for the critical point~\cite{fixed point 4, Phys.Rev.E.100.052146,chenlz}.

The QCD critical point, if exists, is expected to belong to the same universality class of the three-dimensional Ising model~\cite{class 1, class 2, class 3, class 4}. Critical behavior of the corresponding thermodynamics in different systems that belongs to the same universality class is the same which is supervised by the same critical exponents. Recently, many works have been made to map the results of the three-dimensional Ising model to that of the QCD~\cite{PhysRevD102014505, PhysRevC103034901}. Usually, a linear ansatz between the QCD variables, temperature and net-baryon chemical potential, and the Ising variables, temperature and external magnetic field is suggested~\cite{linearmap1, linearmap2, linearmap3, NPA}. Cumulants of net-baryon number, which are the derivatives of the QCD free energy density with respect to net-baryon chemical potential, can be regarded as the combination of the derivatives with respect to temperature and magnetic field in the three-dimensional Ising model in the vicinity of the critical point. Since the critical exponent of external magnetic field is larger than that of the temperature~\cite{Ising exponents1}, the critical behavior of net-baryon number fluctuations is expected to be mainly controlled by the derivatives with respect to the external magnetic field, i.e., the fluctuations of the order parameter in the three-dimensional Ising model.

In this paper, using parametric representation and Monte Carlo simulations of the three-dimensional Ising model, we study and discuss the other kind of fixed point behavior in the temperature dependence of the normalized cumulants and factorial cumulants. Assuming the system formed in the heavy-ion collision experiments is in equilibrium, with a mapping from the Ising model to QCD, the fixed point behavior is also studied and discussed in the energy dependence of the normalized cumulants and factorial cumulants along different freeze-out curves, which may be helpful to locate the QCD critical point.

The paper is organized as follows. In section 2, the three-dimensional Ising model and its parametric representation are introduced. Parametric expressions of second- to fourth-order cumulants and factorial cumulants of the order parameter are derived. At the critical temperature, the independence on the external magnetic fields of the normalized cumulants has been deduced. In section 3, temperature dependence of second- to fourth-order cumulants and factorial cumulants at different distances to the phase boundary and fixed point behavior of the corresponding normalized ones are studied and discussed in the parametric representation. In section 4, the fixed point behavior of normalized second- to fourth-order cumulants and factorial cumulants is discussed in finite-size systems simulated by the Monte Carlo method. In section 5, a mapping from the Ising model to QCD is introduced. The fixed point behavior in the energy dependence of the normalized (factorial) cumulants is studied and discussed. Finally, conclusions and summary are given in section 6.

\section{The second- to fourth-order cumulants and factorial cumulants}

The three-dimensional Ising model is defined as follows,
\begin{flalign}\label{Ising model}
&\qquad \mathcal{H} =-J\sum_{\langle i,j\rangle}{s}_{i}{s}_{j}-H\sum_{i}{s}_{i},&
\end{flalign}
where $\mathcal{H}$ is the Hamiltonian, $s_{i}$ is spin at site $i$ on a simple cubic lattice which can take only two values $\pm 1$. $J$ is the interaction energy between nearest-neighbor spins $\langle i,j\rangle$. $H$ represents the external magnetic field. The magnetization $M$ (the order parameter) is
\begin{flalign}\label{order parameter}
&\qquad M=\frac{1}{V}{\langle \sum_{i}{s_{i}}\rangle}=\frac{\langle s \rangle}{V},&
\end{flalign}
$s=\sum_{i}{s_{i}}$ and $V=L^d$ denotes the total spin and volume of the lattice, respectively, where $d=3$ is the dimension of the lattice and $L$ is the number of lattice points of each direction on the cubic lattice. The magnetization is dependent on the external magnetic field $H$ and the reduced temperature $t=(T-T_c)/{T_c}$, where $T_c$ is the critical temperature. At $t>0$, it is the crossover side. At $t<0$, it is the first order phase transition side.

High-order cumulants of the order parameter can be got from the derivatives of magnetization with respect to $H$ at fixed $t$,
\begin{flalign}\label{cumulants}
&\qquad \left.\kappa_{n}(t,H)=(\frac{\partial^{n-1} M}{\partial H^{n-1}})\right|_{t}.&
\end{flalign}
In particular, the second- to fourth-order cumulants are as follows,
\begin{flalign}\label{second to fourth order cumulant}
&\qquad \kappa_2=\frac{1}{V}{\langle (\delta s)^2 \rangle},& \nonumber \\
&\qquad \kappa_3=\frac{1}{V}{\langle (\delta s)^3 \rangle},& \nonumber \\
&\qquad \kappa_4=\frac{1}{V}(\langle (\delta s)^4 \rangle-3\langle(\delta s)^2 \rangle^2),&
\end{flalign}
where $\delta s=s-\langle s\rangle$.

Turn to the parametric representation of the three-dimensional Ising model, magnetization $M$ and reduced temperature $t$ can be parameterized by two variables $R$ and $\theta$~\cite{linear para, linear para 3},
\begin{flalign}\label{parametric}
&\qquad M=m_0R^{\beta}\theta,~~~~~~t=R(1-\theta^2).&
\end{flalign}
The equation of state of the Ising model can be given by the parametric representation in terms of $R$ and $\theta$ as
\begin{flalign}\label{equation state}
&\qquad H=h_0R^{\beta\delta}h(\theta).&
\end{flalign}
Where $m_0$ in Eq.~\eqref{parametric} and $h_0$ in Eq.~\eqref{equation state} are normalization constants. They are fixed by imposing the normalization conditions $M(t=-1,H=+0)=1$ and $M(t=0,H=1)=1$. $\beta$ and $\delta$ are critical exponents of the three-dimensional Ising universality class with values 0.3267(10) and 4.786(14), respectively~\cite{Ising exponents}.

If $M$, $t$ and $h$ are analytic function of $\theta$, the analytic properties of the equation of state are satisfied~\cite{linear para 1}. The analytic expression of the high-order cumulants can be derived in the parametric representation. What is more, the function $h(\theta)$ is an odd function of $\theta$ because the magnetization is an odd function of the external magnetic field $M(-H)=-M(H)$.

One simple function of $h(\theta)$ obeying all the demands is as follows,
\begin{flalign}\label{equation h}
&\qquad h(\theta)=\theta(3-2\theta^{2}).&
\end{flalign}

This is a mean-field approximation of representation for the equation of state of the three-dimensional Ising model to order $\varepsilon^{2}$, where $\varepsilon$ is a parameter related to the number of dimensions of space. $\varepsilon$-expansion is one of the techniques to explore the critical phenomena. It is enough for our purpose although the parametric representation is also known up to order $\varepsilon^{3}$~\cite{linear para 3}. On the other hand, there is an excellent agreement of the scaling magnetization data from Monte Carlo simulation and the equation of state in the parametric representation~\cite{J. Engels}.

When taking the approximate values of the critical exponents $\beta=1/3$ and $\delta=5$ (it is enough for our purpose), the first fourth-order cumulants in the parametric representation are as follows:
\begin{equation}\label{first four cumulants}
\begin{split}
&\kappa_{1}(t,H)=m_0R^{1/3}\theta,\\
&\kappa_{2}(t,H)=\frac{m_0}{h_0}\frac{1}{R^{4/3}(2\theta^2+3)},\\
&\kappa_{3}(t,H)=\frac{m_0}{h_0^2}\frac{4\theta(\theta^2+9)}{R^{3}(\theta^2-3)(2\theta^2+3)^3},\\
&\kappa_{4}(t,H)=12\frac{m_0}{h_0^3}\frac{(2\theta^8-5\theta^6+105\theta^4-783\theta^2+81)}{R^{14/3}(\theta^2-3)^3(2\theta^2+3)^5}.\\
\end{split}
\end{equation}

The reduced temperature $t$ and external magnetic field $H$, are functions of $R$ and $\theta$ provided by Eq.~\eqref{parametric} and Eq.~\eqref{equation state}. At fixed $H$, $R$ can be represented in terms of $\theta$ by Eq.~\eqref{equation state}. As a consequence, cumulants in Eq.~\eqref{first four cumulants} just depends on $\theta$, so is the reduced temperature $t$ in Eq.\eqref{parametric}. There are three kinds of special values of $\theta$, they are $\theta=\theta_{n}^{max}$ for the peak of $\kappa_n$ if the peak exists, $\theta=\theta_{n}^{min}$ for the valley of $\kappa_n$ if the valley exists, $\theta=1$ for the reduced temperature $t=0$ (the critical temperature) at a positive magnetic field (or $\theta=-1$ for $t=0$ at a negative magnetic field), respectively.

The first two cases imply the ratios (the factor of $H$ is offset in the ratios) of the peak hight to the valley depth for $\kappa_4$, $\kappa_5$ and $\kappa_6$ are universal and independent of $H$. They are approximately $-28$, $-0.1$, and $-6$ for $H>0$, respectively~\cite{Stephanov-prl107, panx}.

At a positive magnetic field, temperature dependence of even-order cumulants shows a positive peak in the vicinity of the critical temperature, while it is a negative valley for the odd-order cumulants~\cite{panx}.
Normalizing the even-order cumulants by their peak hight $\kappa_{2n}^{max}, n=1,2,3...$, and the odd-order cumulants by the absolute value of the valley depth $\vert\kappa_{2n+1}^{min}\vert, n=1,2,3...$, then from the last case, one can get a fixed point behavior of temperature dependence of normalized cumulants $\kappa_n^{Norm}$ for different values of $H$ at $t=0$.

\begin{figure*}[hbt]
\centering
    \includegraphics[width=0.3\textwidth]{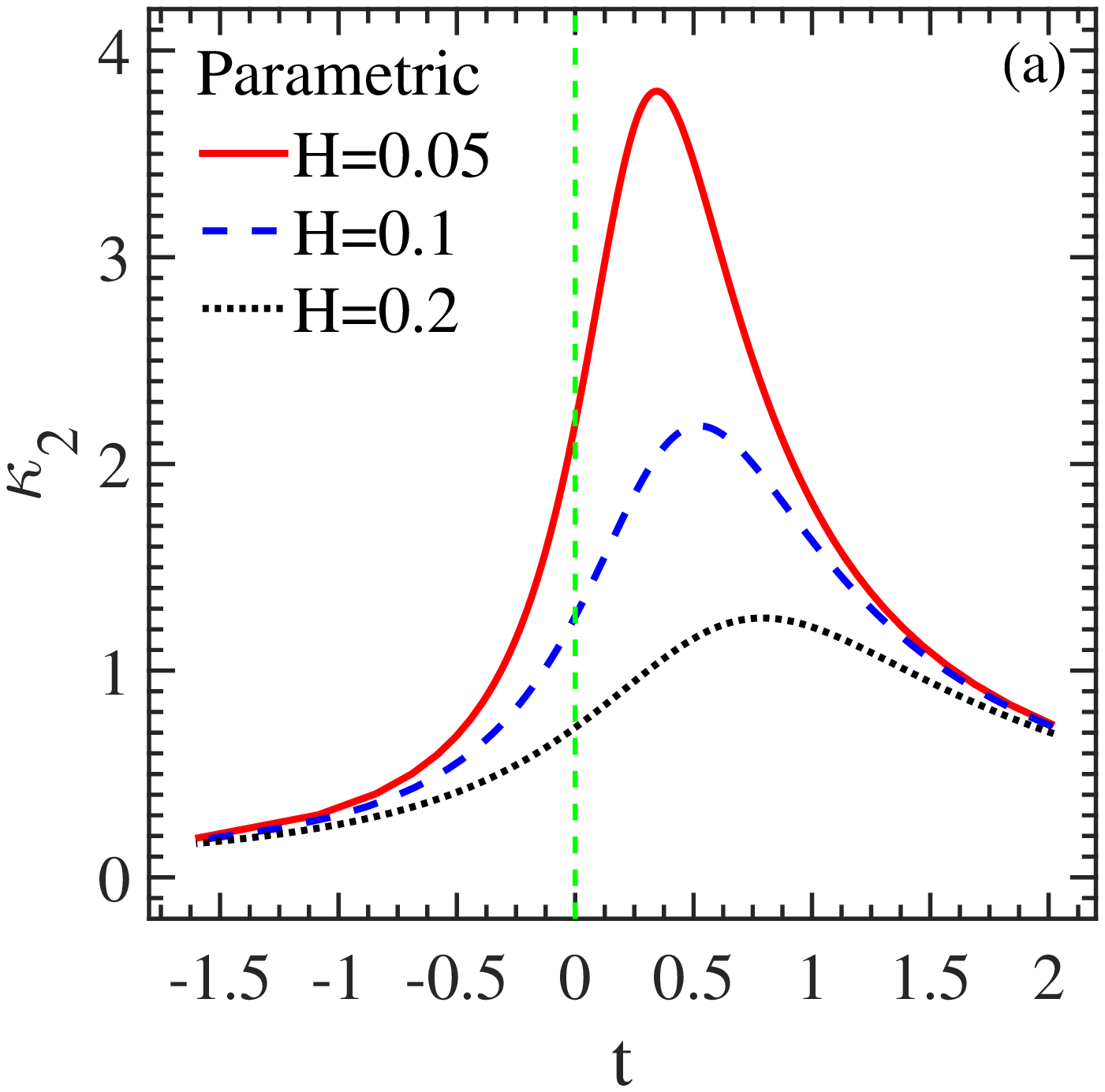}
    \includegraphics[width=0.3\textwidth]{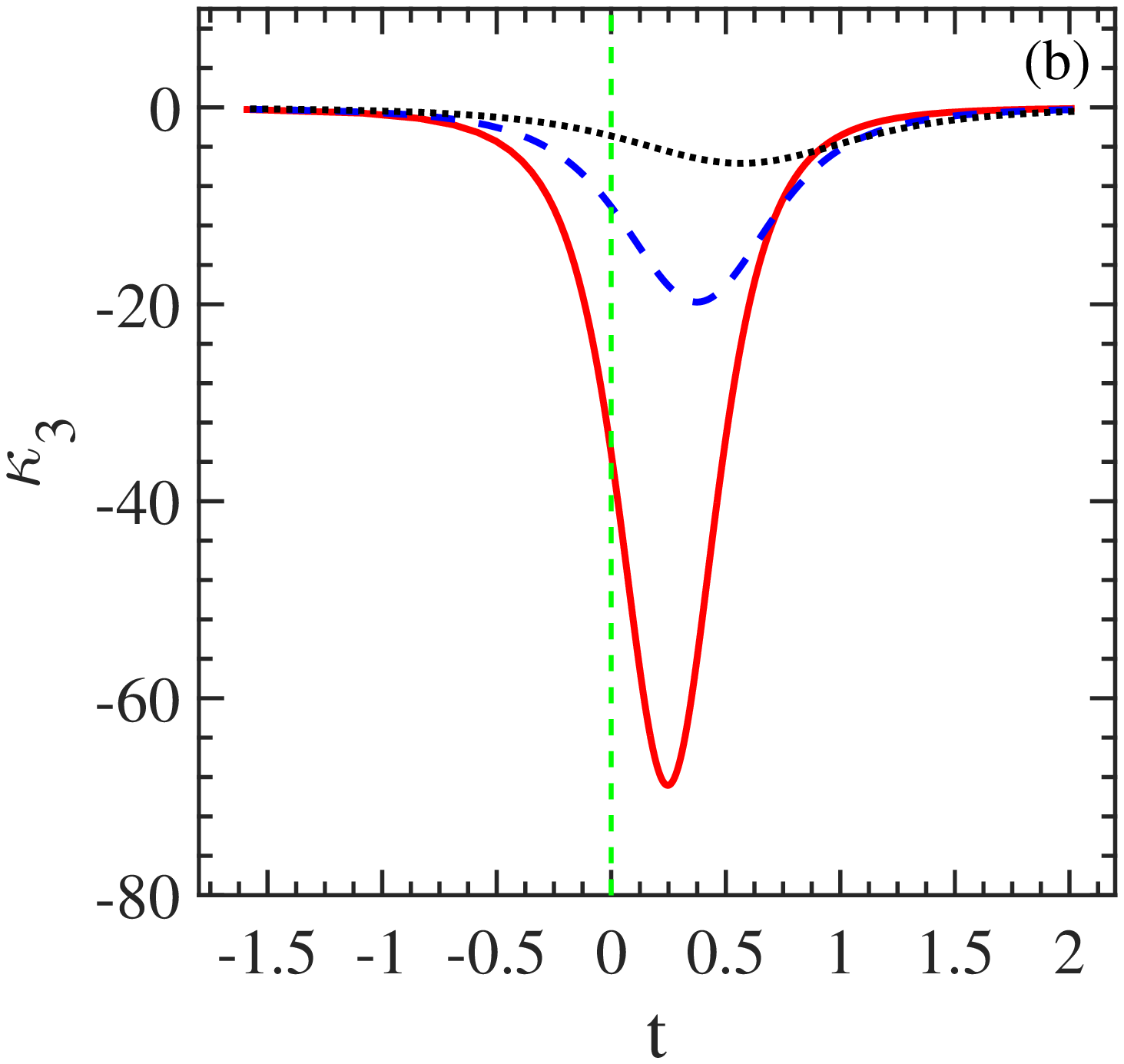}
    \includegraphics[width=0.3\textwidth]{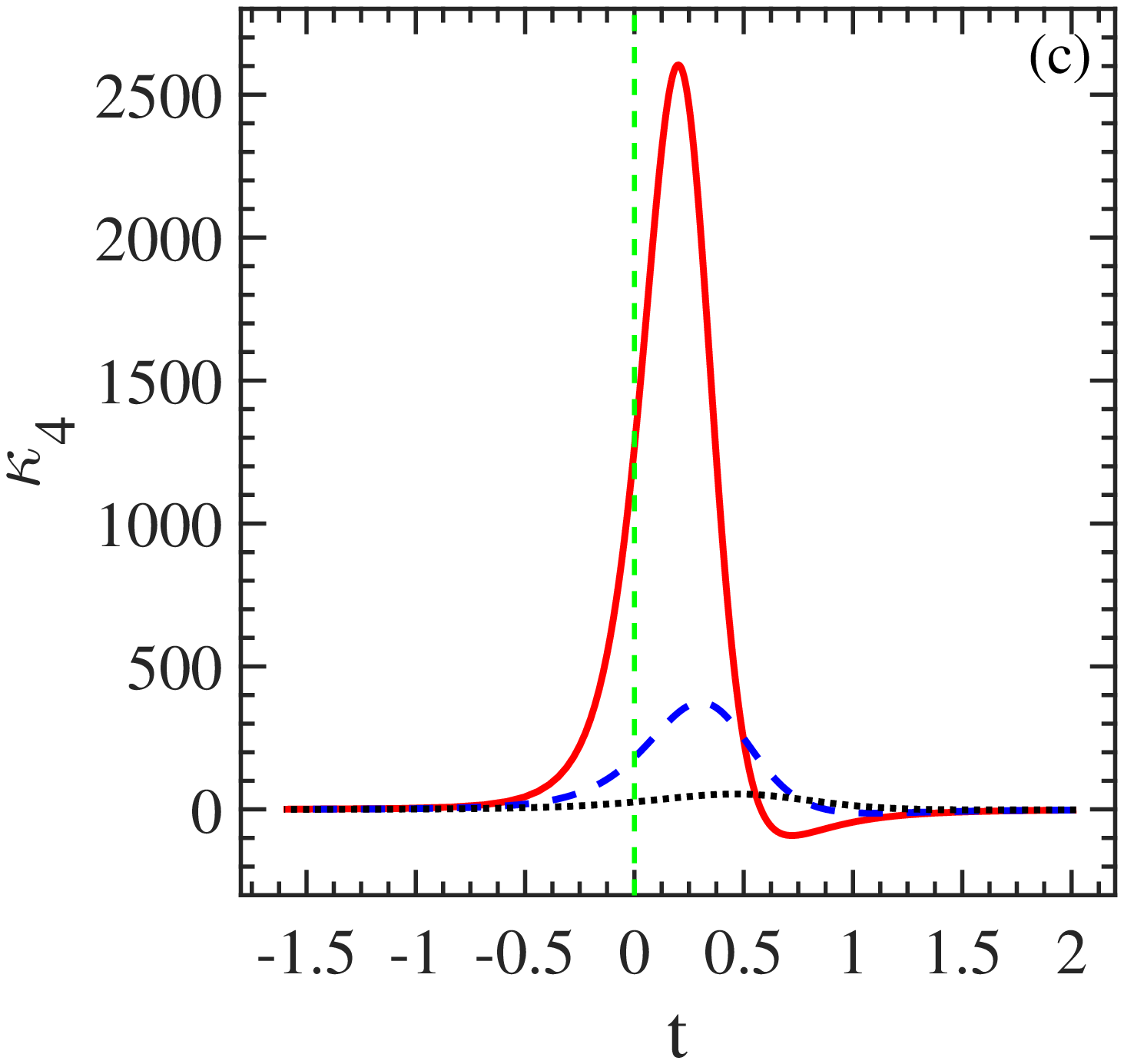}
    \includegraphics[width=0.3\textwidth]{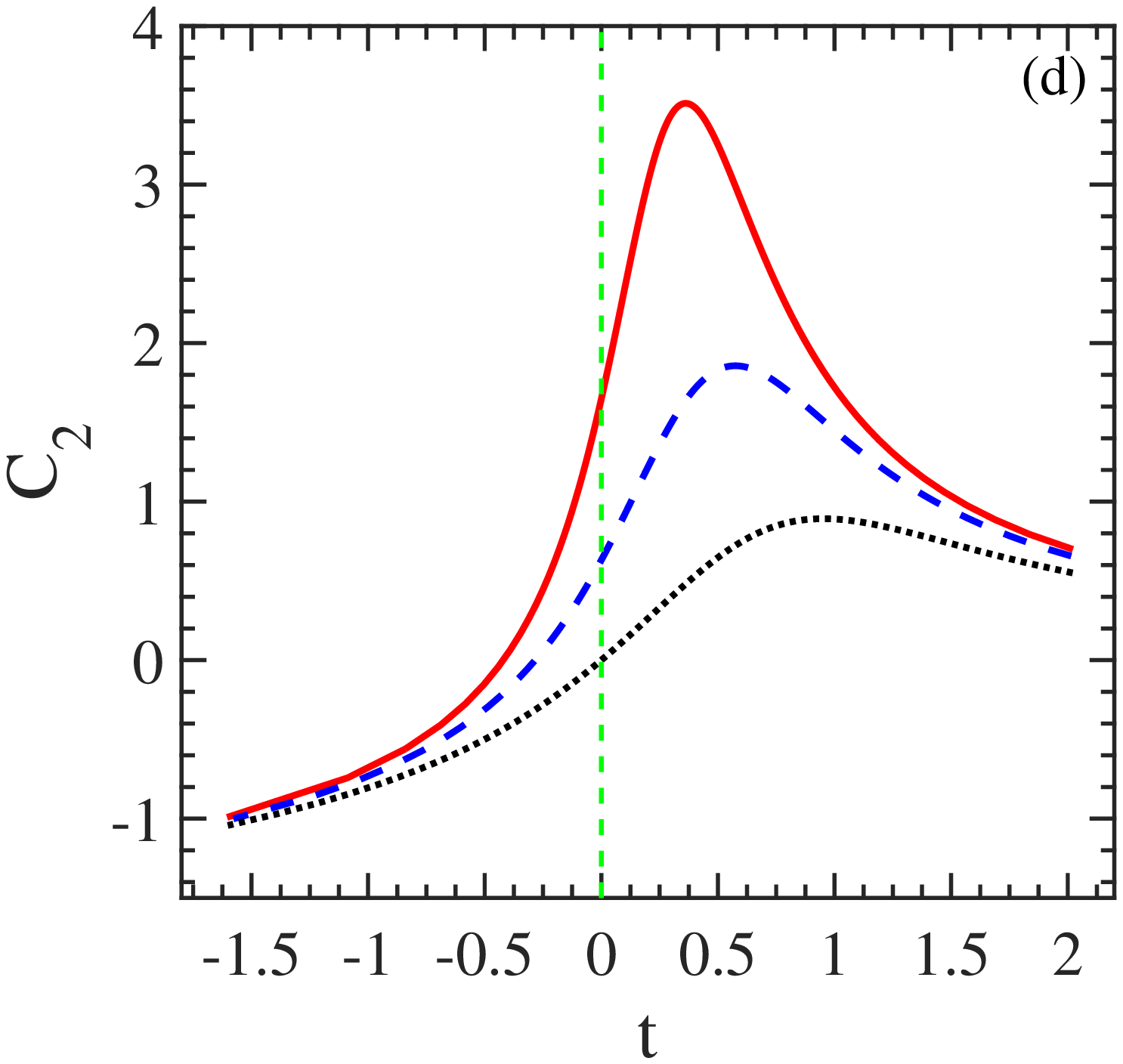}
    \includegraphics[width=0.3\textwidth]{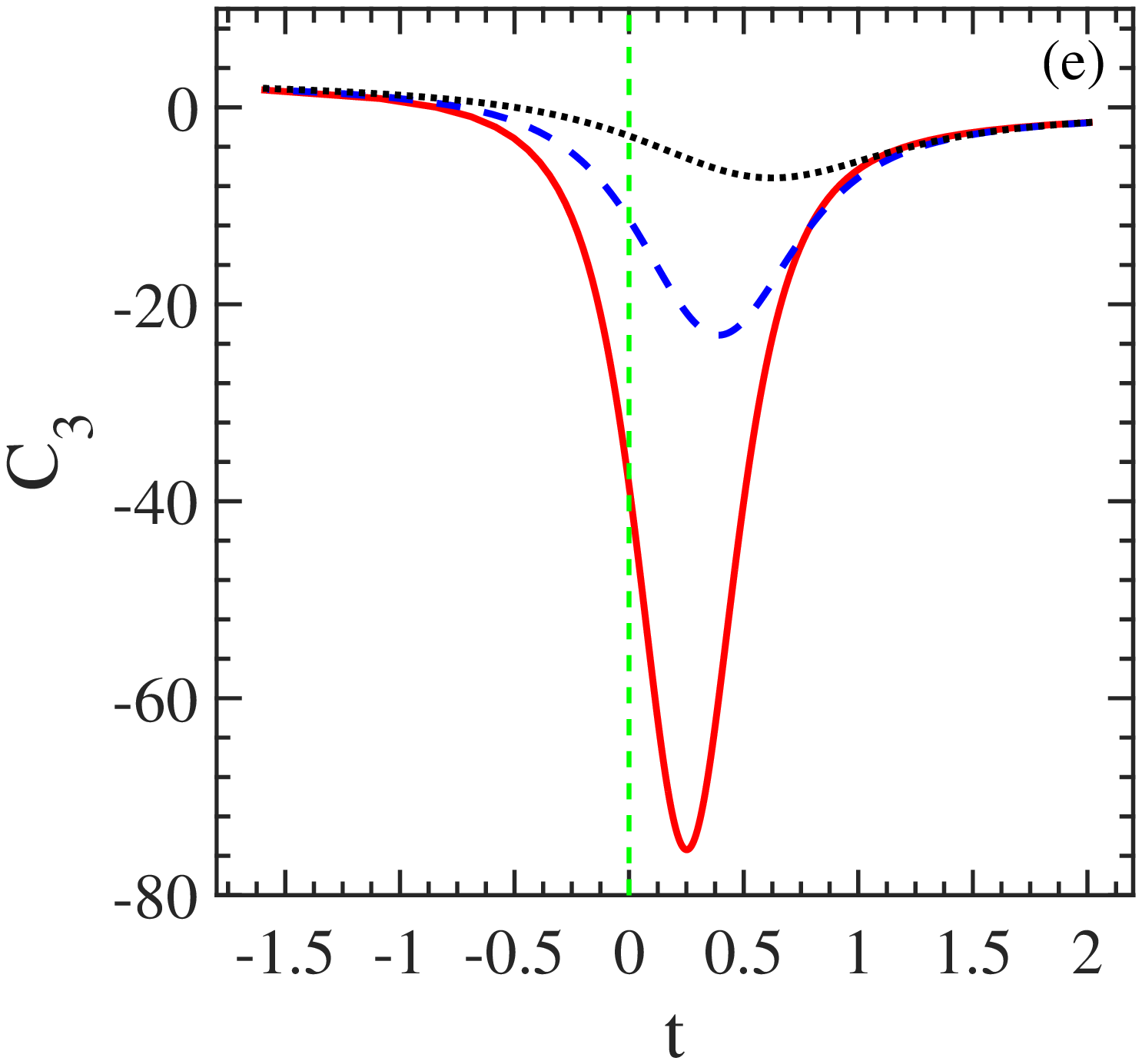}
    \includegraphics[width=0.3\textwidth]{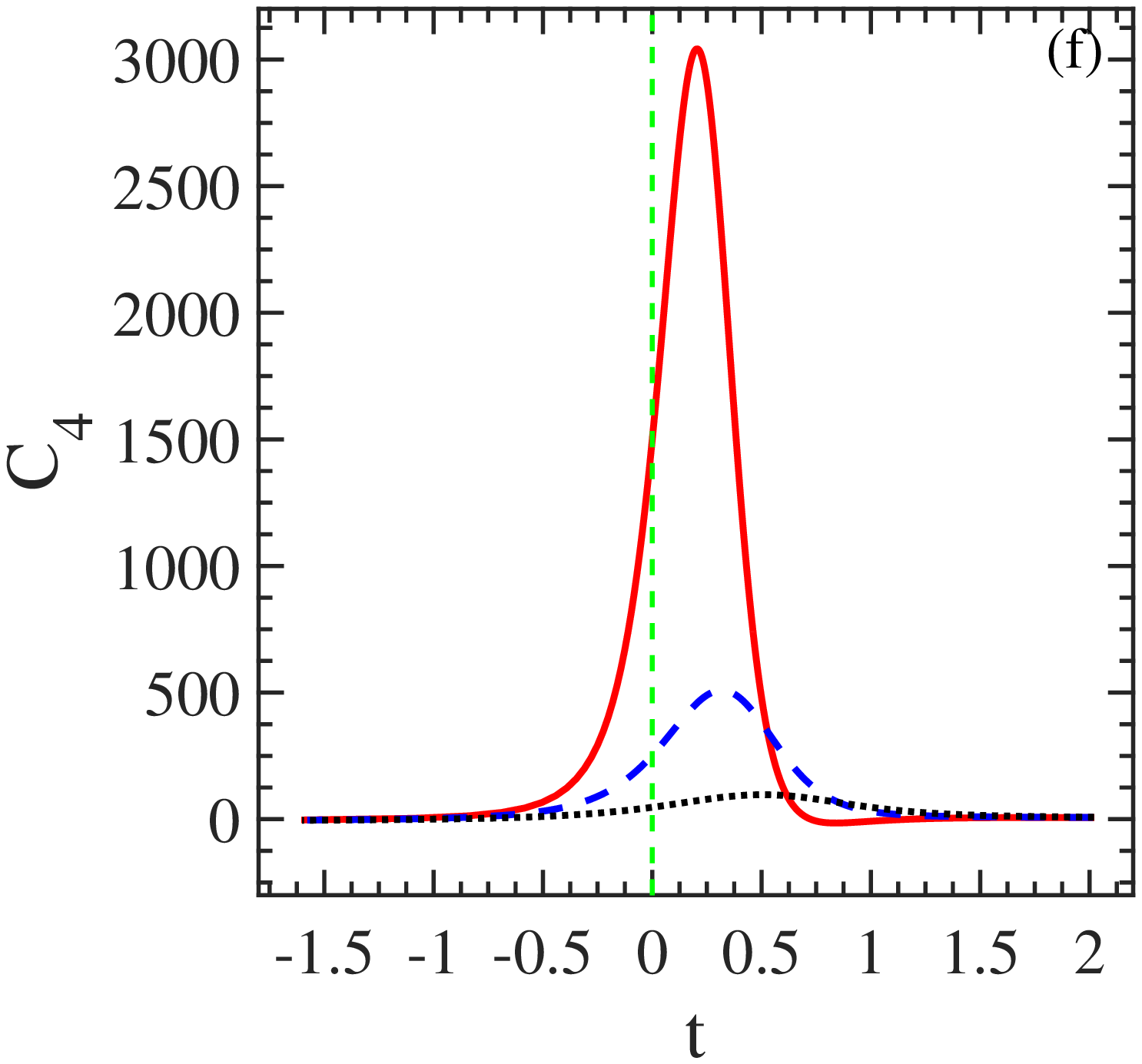}
    \caption{\label{Fig. 1}(Color online). Temperature dependence of $\kappa_2$ (a), $\kappa_3$ (b), $\kappa_4$ (c), $C_2$ (d), $C_3$ (e) and $C_4$ (f) at three different values of external magnetic fields, $H = 0.05$, $0.1$ and $0.2$, in the parametric representation of the three-dimensional Ising model. The green dashed line shows the critical temperature.}
\end{figure*}

Especially, one can get the second- to fourth-order normalized cumulants,
\begin{equation}\label{normalized cumulants}
\begin{split}
&\kappa_{2}^{Norm}=\kappa_{2}/\kappa_{2}^{max},\\
&\kappa_{3}^{Norm}=\kappa_{3}/\vert\kappa_{3}^{min}\vert, \\
&\kappa_{4}^{Norm}=\kappa_{4}/\kappa_{4}^{max}.\\
\end{split}
\end{equation}

At any positive magnetic field, values of second- to fourth-order normalized cumulants at $t=0$ are as follows,
\begin{equation}\label{normalized cumulants at t=0}
\begin{split}
&\kappa_{2}^{Norm}(t=0)=\frac{\kappa_{2}(\theta=1)}{\kappa_{2}(\theta=\theta_2^{max})}\approx 0.58, \\
&\kappa_{3}^{Norm}(t=0)=\frac{\kappa_{3}(\theta=1)}{\vert \kappa_{3}(\theta=\theta_3^{min})\vert}\approx -0.51,\\
&\kappa_{4}^{Norm}(t=0)=\frac{\kappa_{4}(\theta=1)}{\kappa_{4}(\theta=\theta_4^{max})}\approx 0.49. \\
\end{split}
\end{equation}

In fact, cumulants can be normalized by their values at any $\theta$ to get the fixed point behavior at the critical temperature, but among those the most convenient choice would be normalization by the extreme values which can be identified easily from measured data.

The second- to fourth-order factorial cumulants can be expressed by the cumulants as follows~\cite{Phys. Rev. C.96.024910},
\begin{equation}\label{first four order factorial cumulants}
\begin{split}
&C_{2}=\kappa_{2}-\kappa_{1},\\
&C_{3}=\kappa_{3}-3\kappa_{2}+2\kappa_{1},\\
&C_{4}=\kappa_{4}-6\kappa_{3}+11\kappa_{2}-6\kappa_{1}.\\
\end{split}
\end{equation}

They can also be normalized by their maximum or the absolute values of their minimum as follows,
\begin{equation}\label{normalized factorial cumulants}
\begin{split}
&C_{2}^{Norm}=C_{2}/C_{2}^{max},\\
&C_{3}^{Norm}=C_{3}/\vert C_{3}^{min}\vert, \\
&C_{4}^{Norm}=C_{4}/C_{4}^{max}.\\
\end{split}
\end{equation}

Because the factorial cumulants mix different orders of cumulants as showed in Eq.~\eqref{first four order factorial cumulants}, far away from the critical point, the behavior of factorial cumulants is very different from the same order cumulants~\cite{panx}. It also appears to be that there may be no fixed point behavior in the temperature dependence of the factorial cumulants at the critical temperature for different external magnetic fields. But one should keep in mind that in the vicinity of the critical point, cumulants and the same order factorial cumulants can not be distinguished. The higher the order of the factorial cumulant, the more dominant role of the same order cumulant in its critical behavior.

\section{Fixed point behavior of normalized cumulants and factorial cumulants in the parametric representation}

As the increase of value of $H$, it is far away from the phase boundary. At three different magnetic fields $H=0.05, 0.1, 0.2$, the temperature dependence of second- to fourth-order cumulants and factorial cumulants are studied in the parametric representation of the three-dimensional Ising model, as showed in Fig.~1(a) to 1(f). The vertical green dashed line shows the critical temperature.

It is clear that, as the decreasing value of $H$, the qualitative temperature dependence of $\kappa_2$ does not change, all showing a peak structure in the vicinity of the critical temperature. But the peak becomes higher, sharper and closer to the critical temperature as showed in Fig.~1(a). The similar situation occurs for $\kappa_3$ in Fig.~1(b) and $\kappa_4$ in Fig.~1(c). The smaller value of $H$, the closer to the phase boundary, the more singular of the behavior of cumulants.

In the vicinity of the critical temperature, trends of temperature dependence of factorial cumulants are similar with the same order cumulants as showed in Fig.~1(d) to 1(f). When it is far away from the critical temperature, the sign of factorial cumulants is possible to change, this is consistent with the results in Ref.~\cite{panx}.

The normalized cumulants and factorial cumulants are showed in Fig.~2. The vertical green dashed line shows the critical temperature, while the horizontal green dashed line shows the value of normalized cumulants at the critical temperature which inferred from Eq.~\eqref{normalized cumulants at t=0}.

\begin{figure*}[hbt]
\centering
    \includegraphics[width=0.3\textwidth]{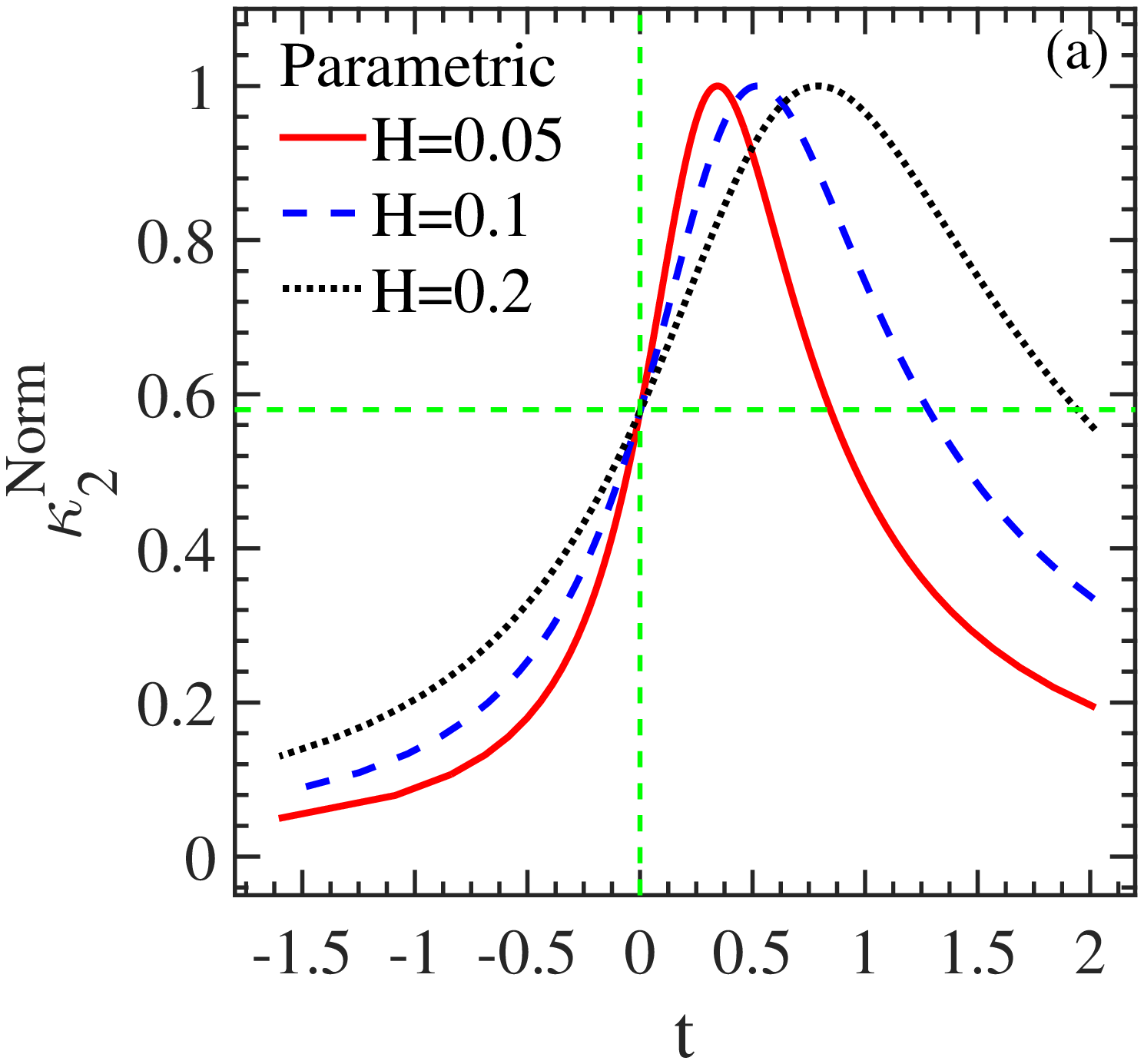}
    \includegraphics[width=0.3\textwidth]{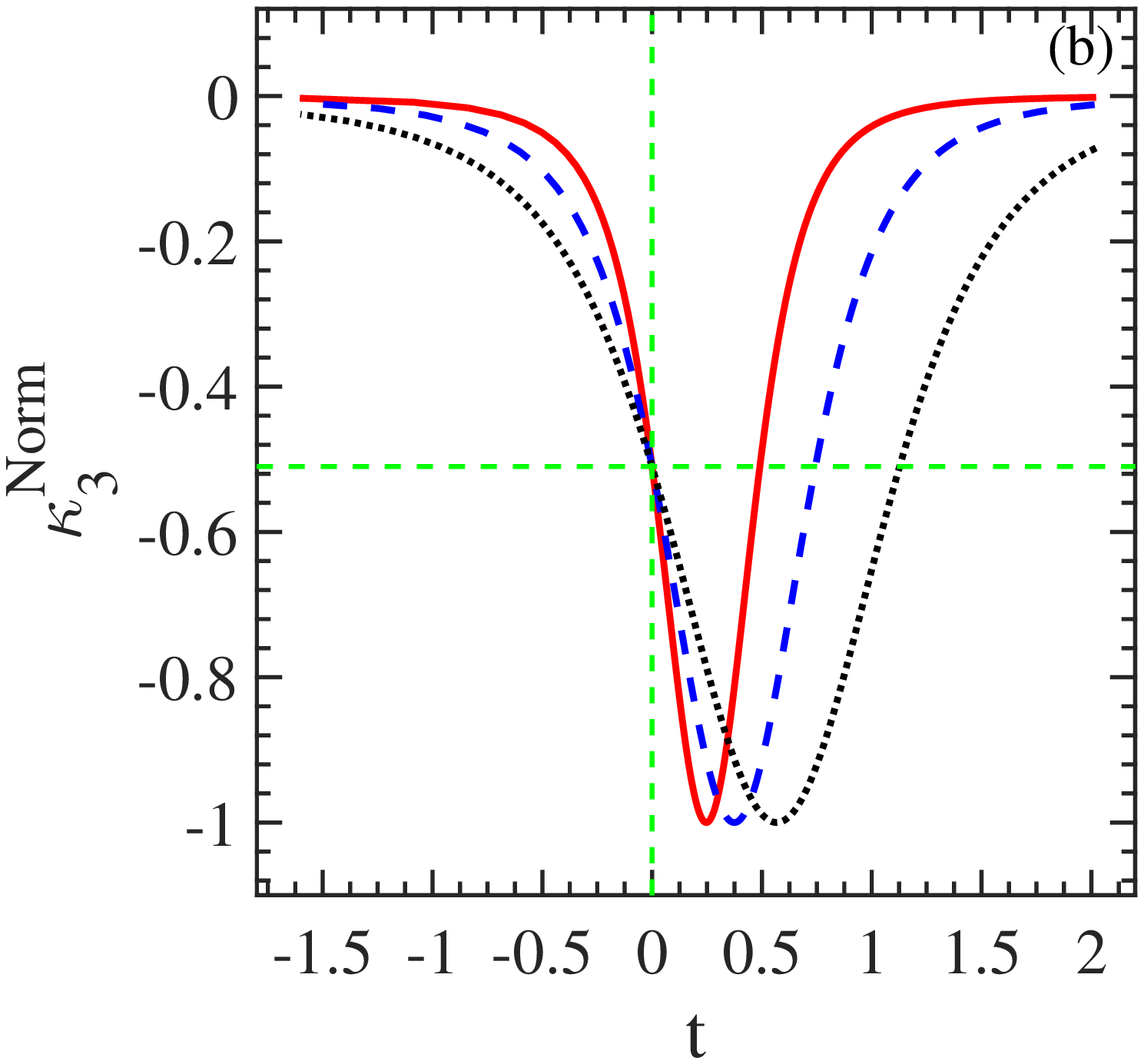}
    \includegraphics[width=0.3\textwidth]{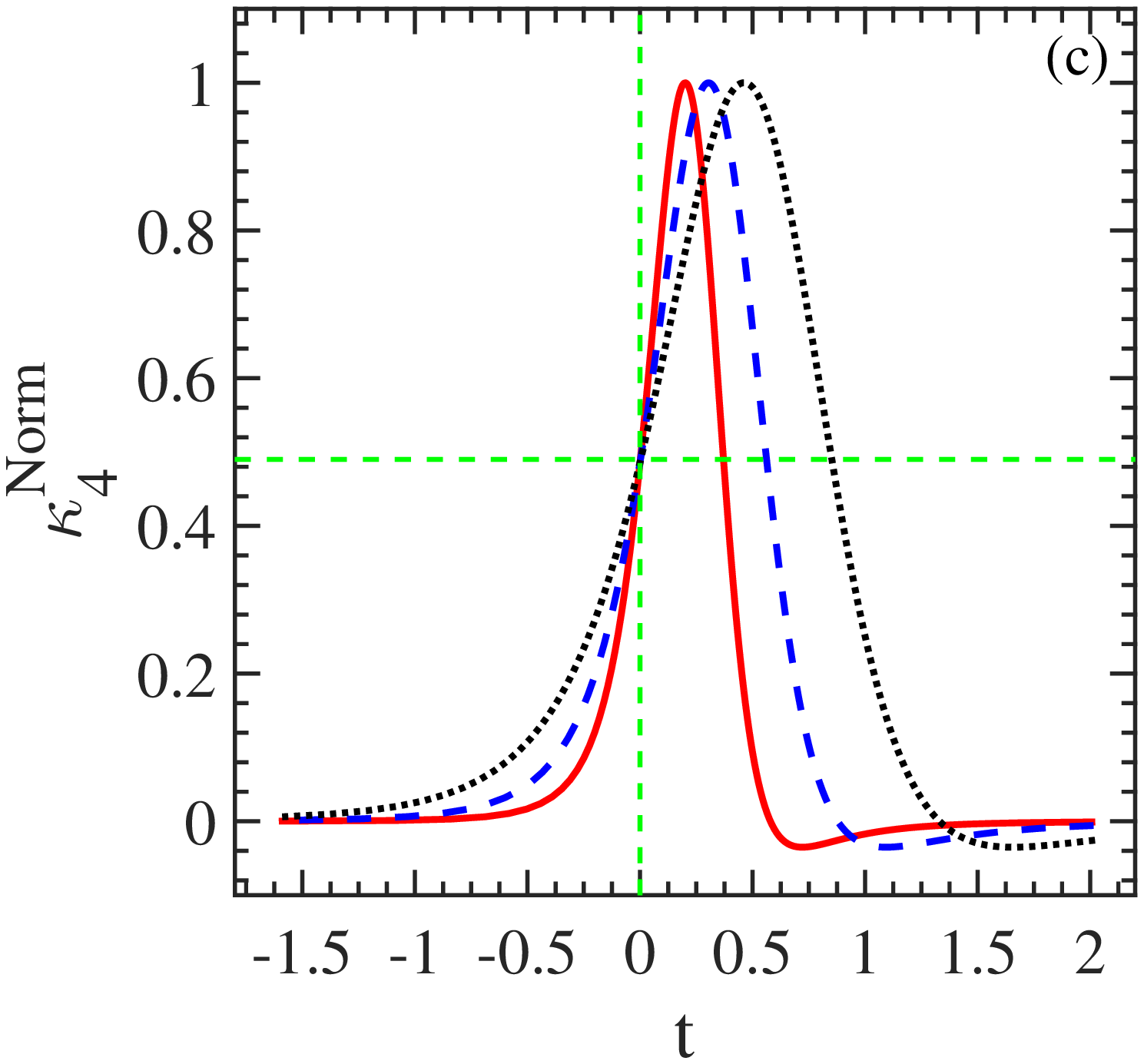}
    \includegraphics[width=0.3\textwidth]{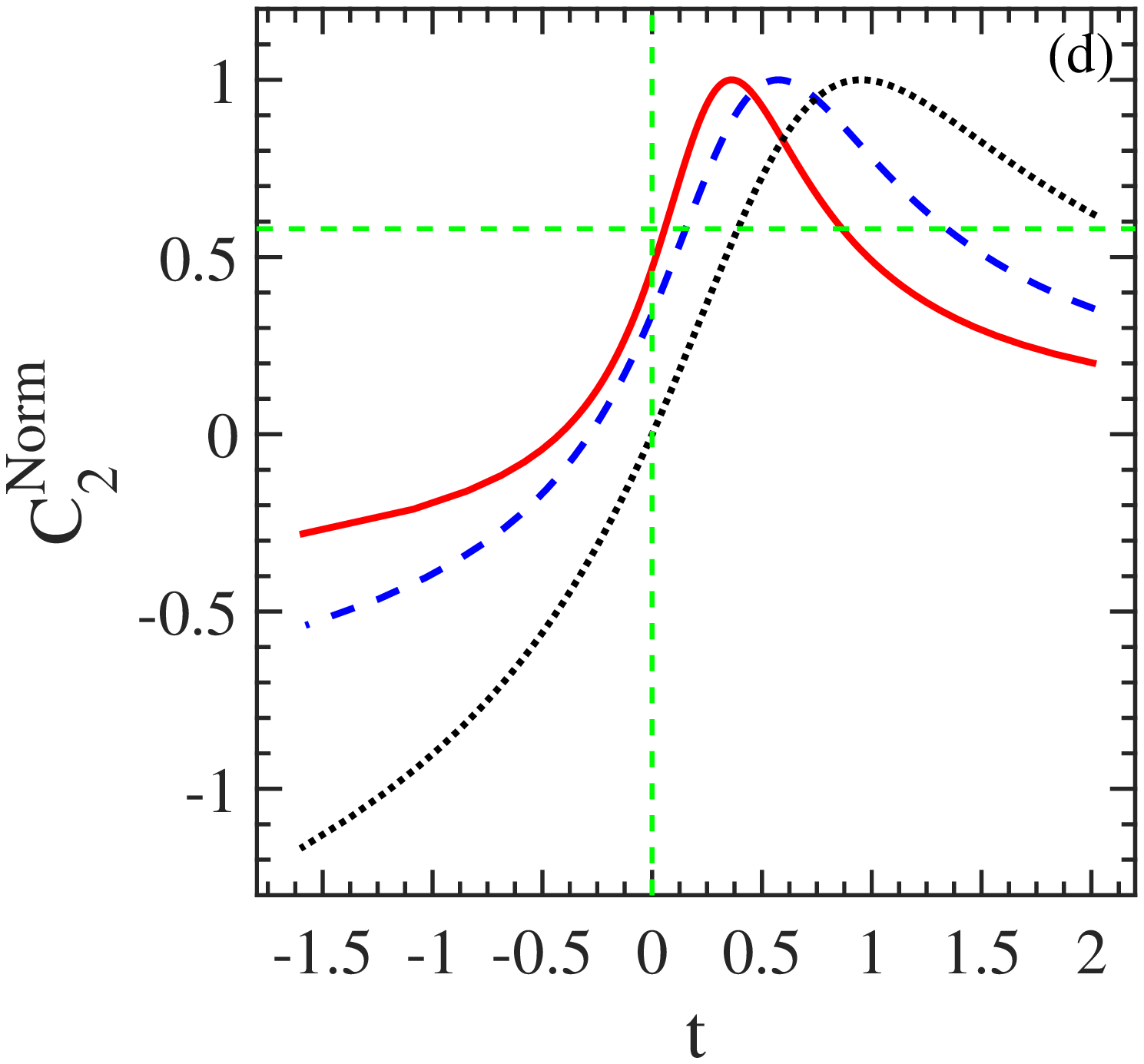}
    \includegraphics[width=0.3\textwidth]{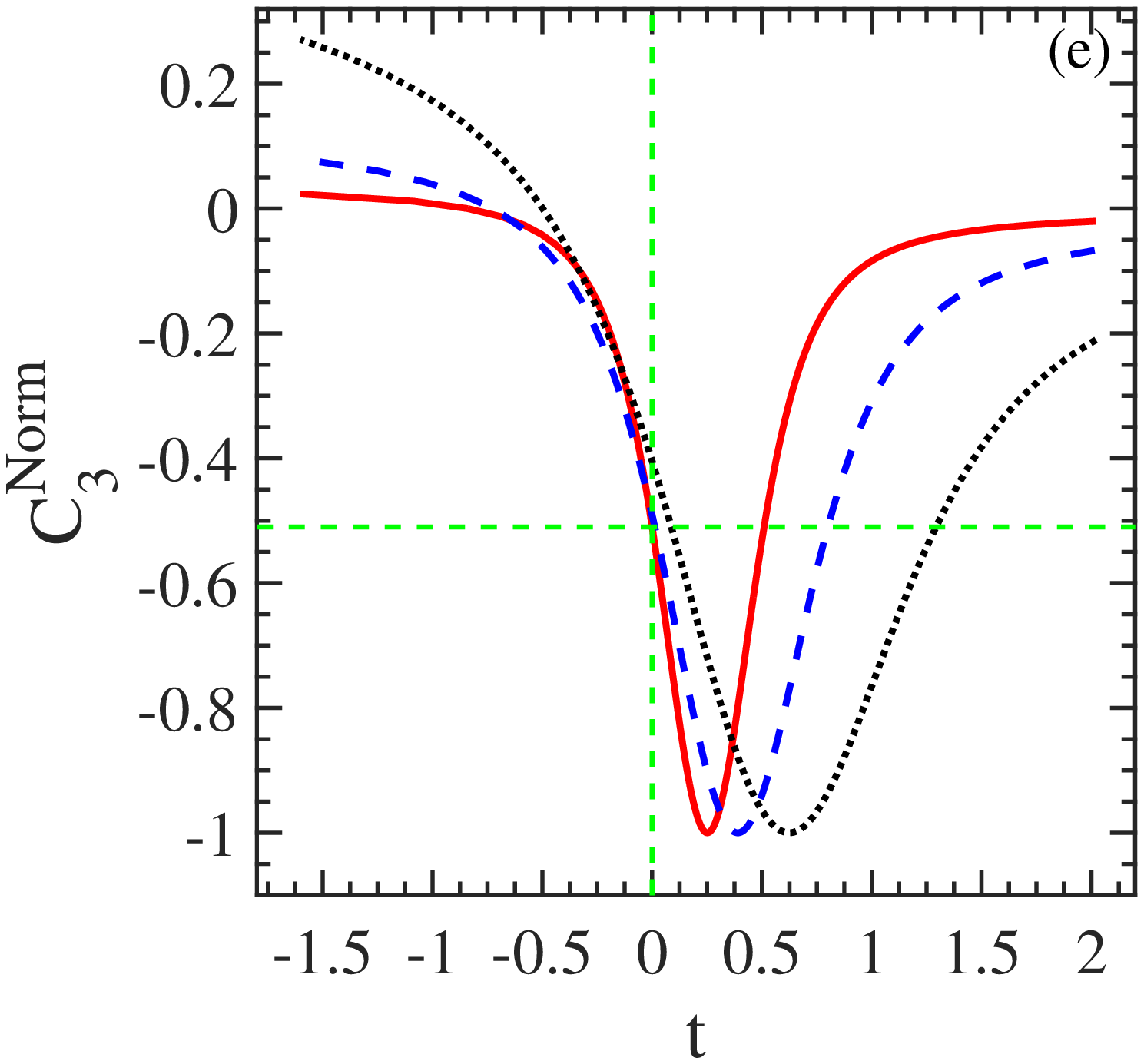}
    \includegraphics[width=0.3\textwidth]{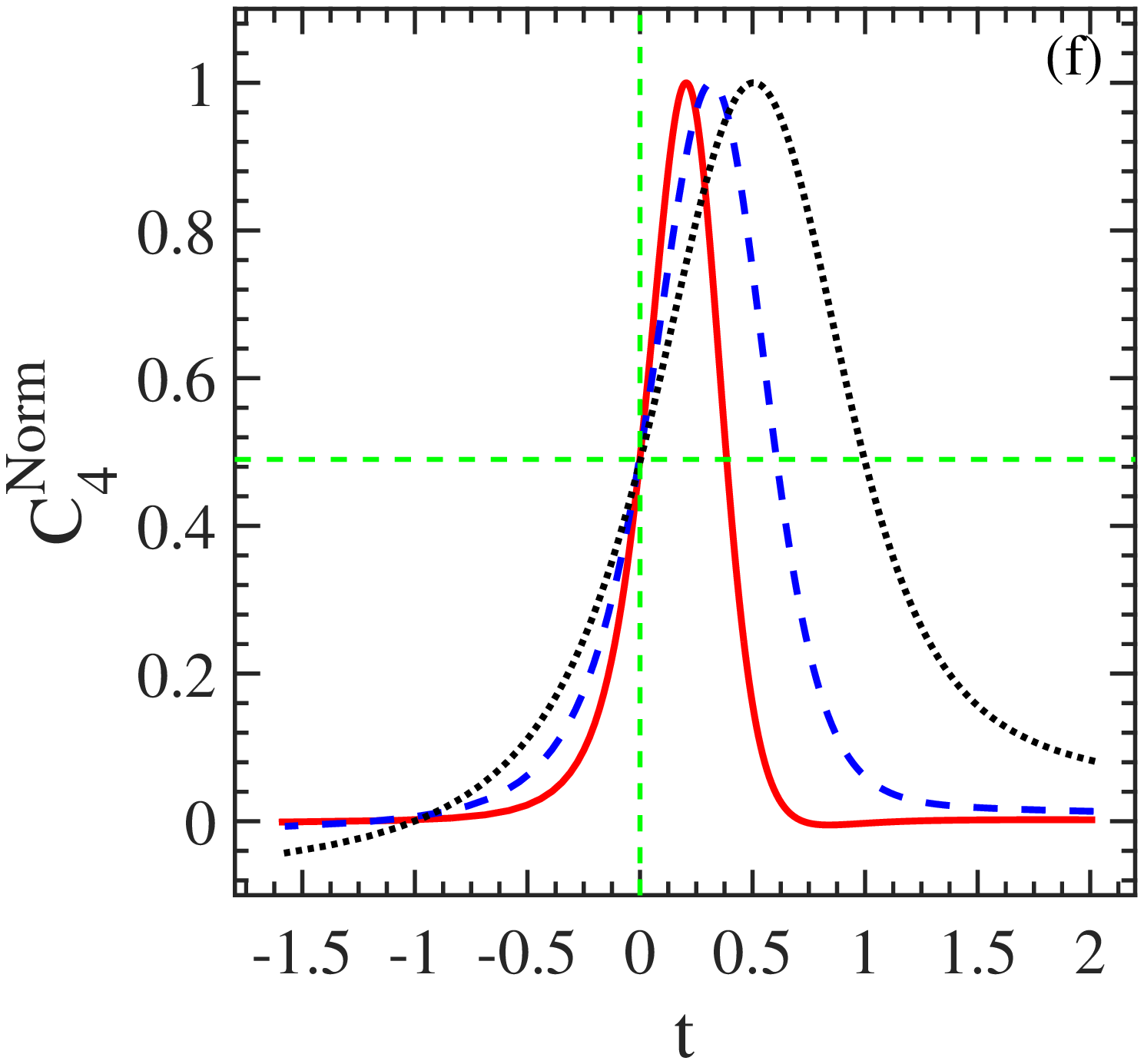}
    \caption{\label{Fig. 3}(Color online). Temperature dependence of $\kappa_2^{Norm}$ (a), $\kappa_3^{Norm}$ (b), $\kappa_4^{Norm}$ (c), $C_2^{Norm}$ (d), $C_3^{Norm}$ (e) and $C_4^{Norm}$ (f) at three different values of external magnetic fields, $H = 0.05$, $0.1$ and $0.2$, in the parametric representation of the three-dimensional Ising model. The cross point of green dashed lines is the fixed point.}
\end{figure*}

It is clear that for $\kappa_2^{Norm}$, $\kappa_3^{Norm}$ and $\kappa_4^{Norm}$ showed in Fig.~2(a) to 2(c), a common feature occurs. That is the fixed point behavior at the critical temperature. At different values of $H$, values of the $\kappa_2^{Norm}$ are the same at the critical temperature. It is independent of the distance to the phase boundary. So are the values of $\kappa_3^{Norm}$ and $\kappa_4^{Norm}$. The fixed point is just at the cross point of the two dashed green lines. That is to say the values of the normalized cumulants at $t$ are consistent with Eq.~\eqref{normalized cumulants at t=0}.

As shown in Fig.~2(c), the valley depths for $\kappa_4^{Norm}$ at $H = 0.05, 0.1, 0.2$ are almost the same~\cite{Stephanov-prl107}. One can easily conclude that the ratios of the peak hight to the valley depth are independent of $H$. For the fourth-order cumulant, in some cases, if the peak can not be decided, one can normalize it by its valley depth. The fixed point behavior also exists.

All in all, the ratios of the value of even-order cumulants (odd-order cumulants) at critical temperature to its peak value (valley depth) is independent on the external magnetic fields. This results in a fixed point behavior in the temperature dependence of normalized cumulants, which may be helpful to search for the critical temperature.

Turn to the normalized factorial cumulants showed in Fig.~2(d) to 2(f), it is clear that far away from the critical temperature, each order of factorial cumulant has sign changes as the increasing $H$. That is to say far away from the phase boundary, there exists sign difference between cumulants and the same order factorial cumulants.

Let us pay attention to the fixed point behavior of the normalized factorial cumulants, there is no fixed point behavior in temperature dependence of $C_2^{Norm}$ as showed in Fig.~2(d), which is in line with the inference from relation of factorial cumulants with cumulants in Eq.~\eqref{first four order factorial cumulants}.

For $C_3^{Norm}$ in Fig.~2(e), the fixed point behavior is not so obvious. But the fixed point occurs again in $C_4^{Norm}$ as showed in Fig.~2(f). The position of the fixed point is just at the cross point of the two green dashed line, consistent with $\kappa_4^{Norm}$. In fact, the higher the order of the cumulants, the more sensitive of the cumulants to the correlation length, the more dominant role of the cumulants in the critical behavior of the same order factorial cumulants. So fixed point behavior occurring in $C_4^{Norm}$ again is not hard to understand.

\section{Fixed point behavior of normalized cumulants and factorial cumulants by Monte Carlo simulations}

\begin{figure*}[hbt]
\centering
    \includegraphics[width=0.3\textwidth]{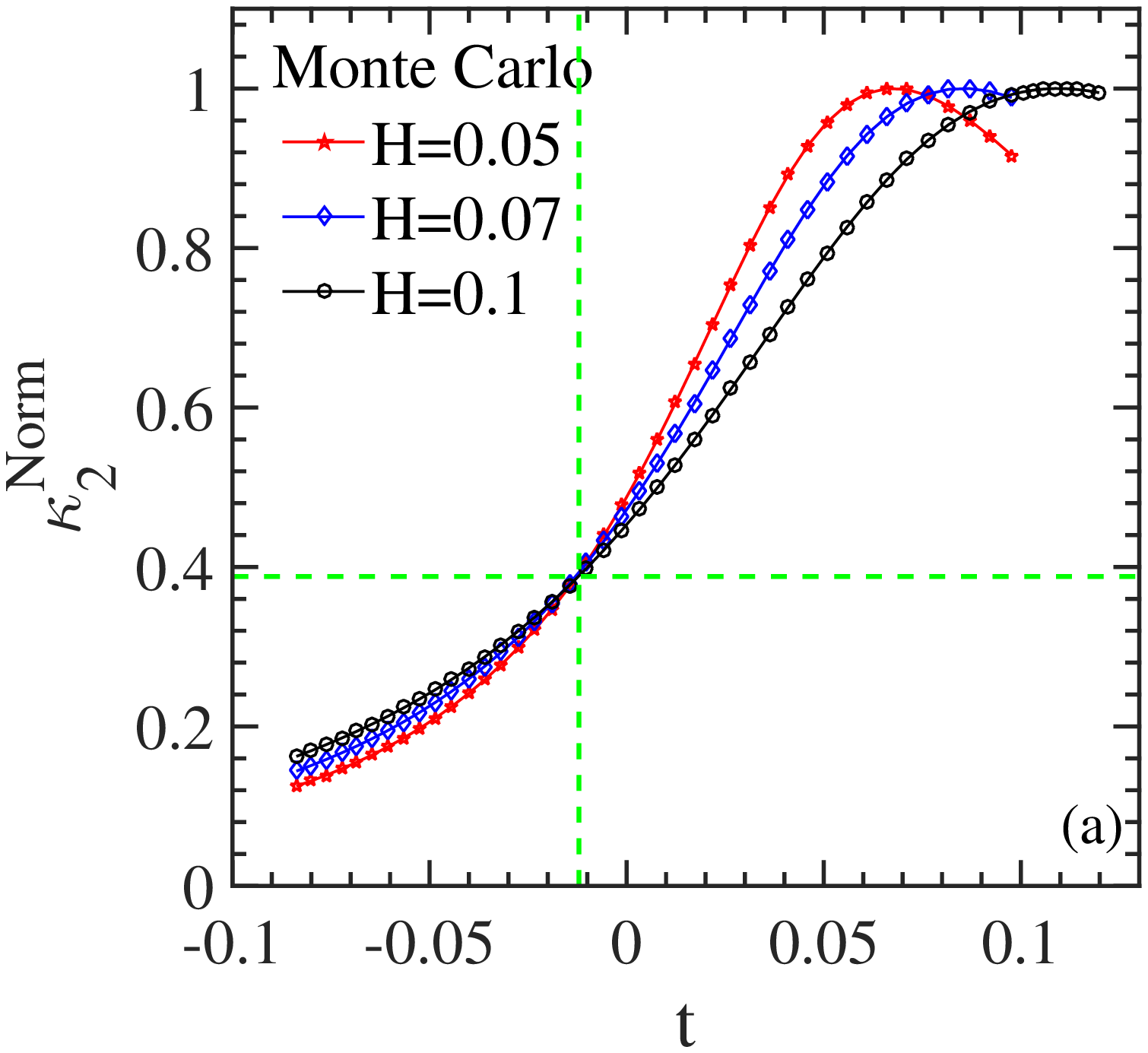}
    \includegraphics[width=0.3\textwidth]{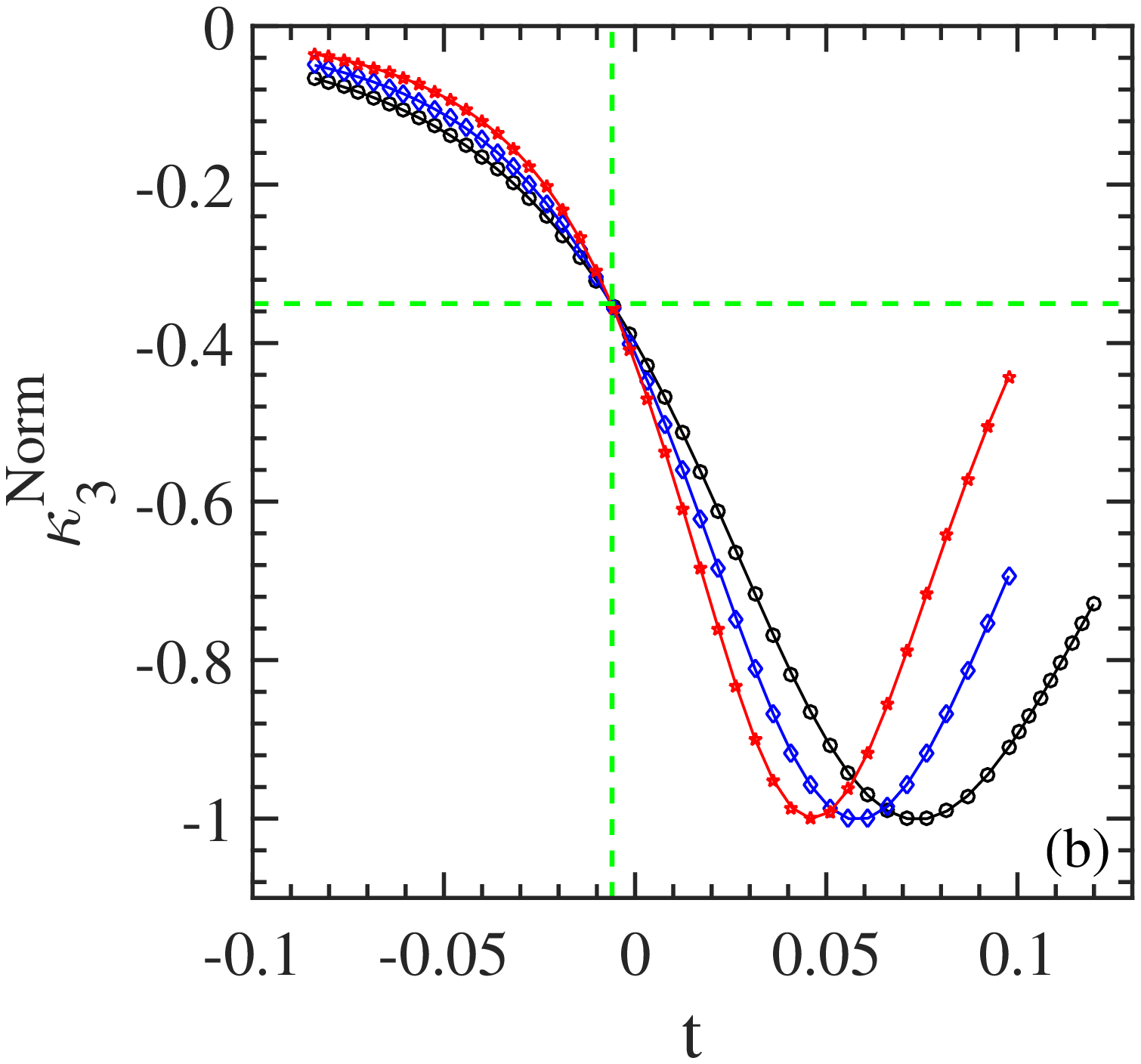}
    \includegraphics[width=0.3\textwidth]{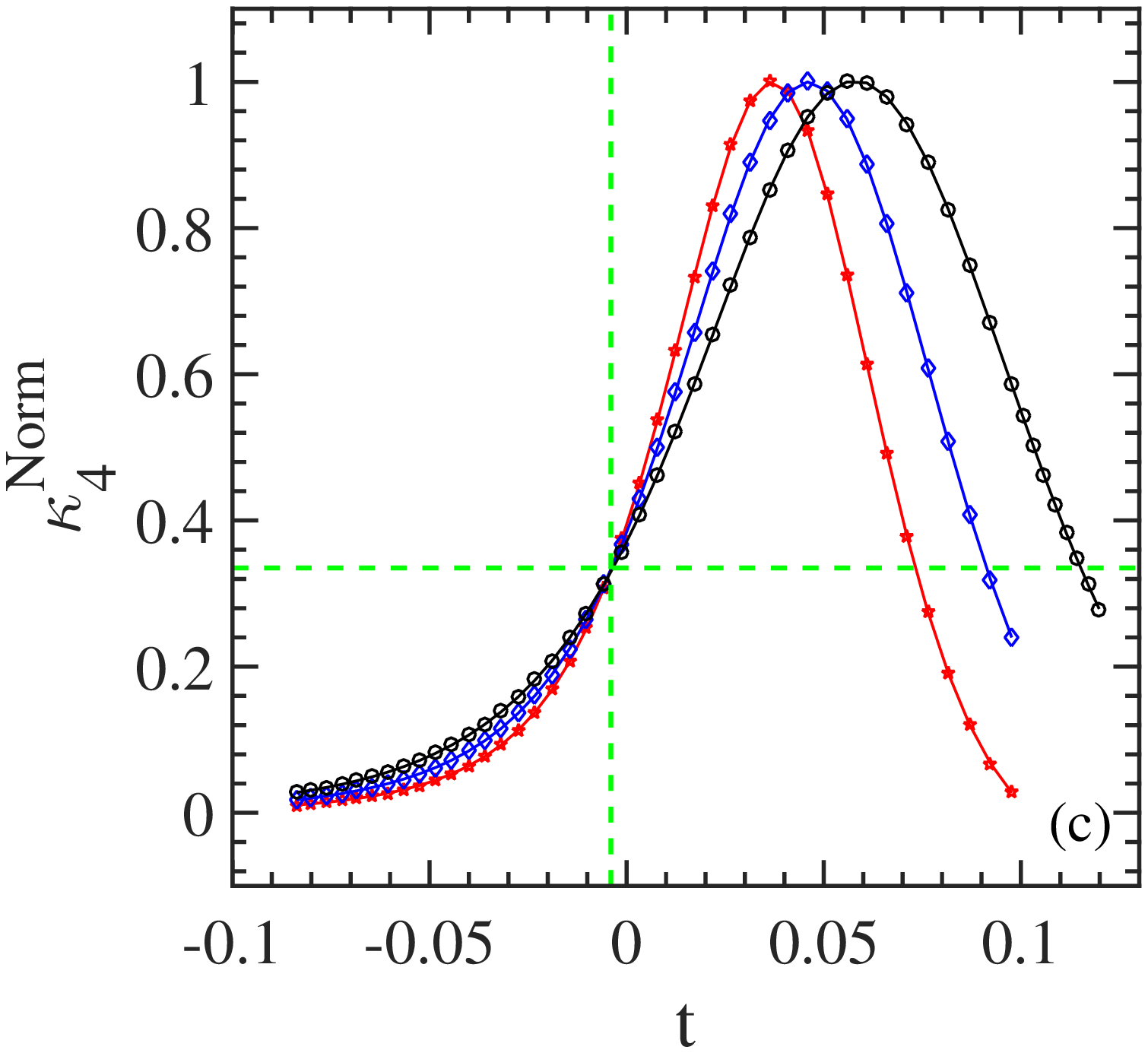}
     \includegraphics[width=0.3\textwidth]{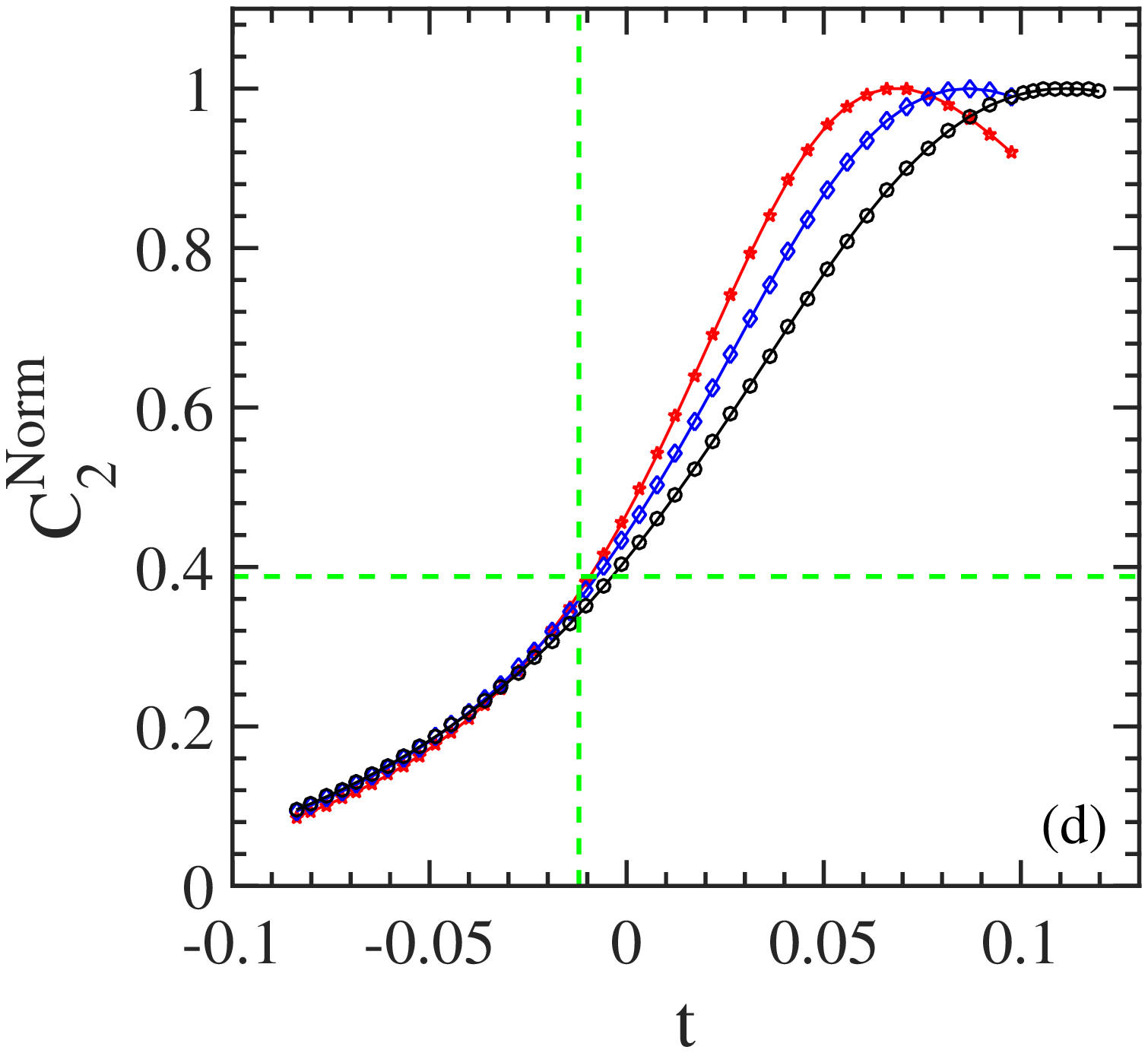}
    \includegraphics[width=0.3\textwidth]{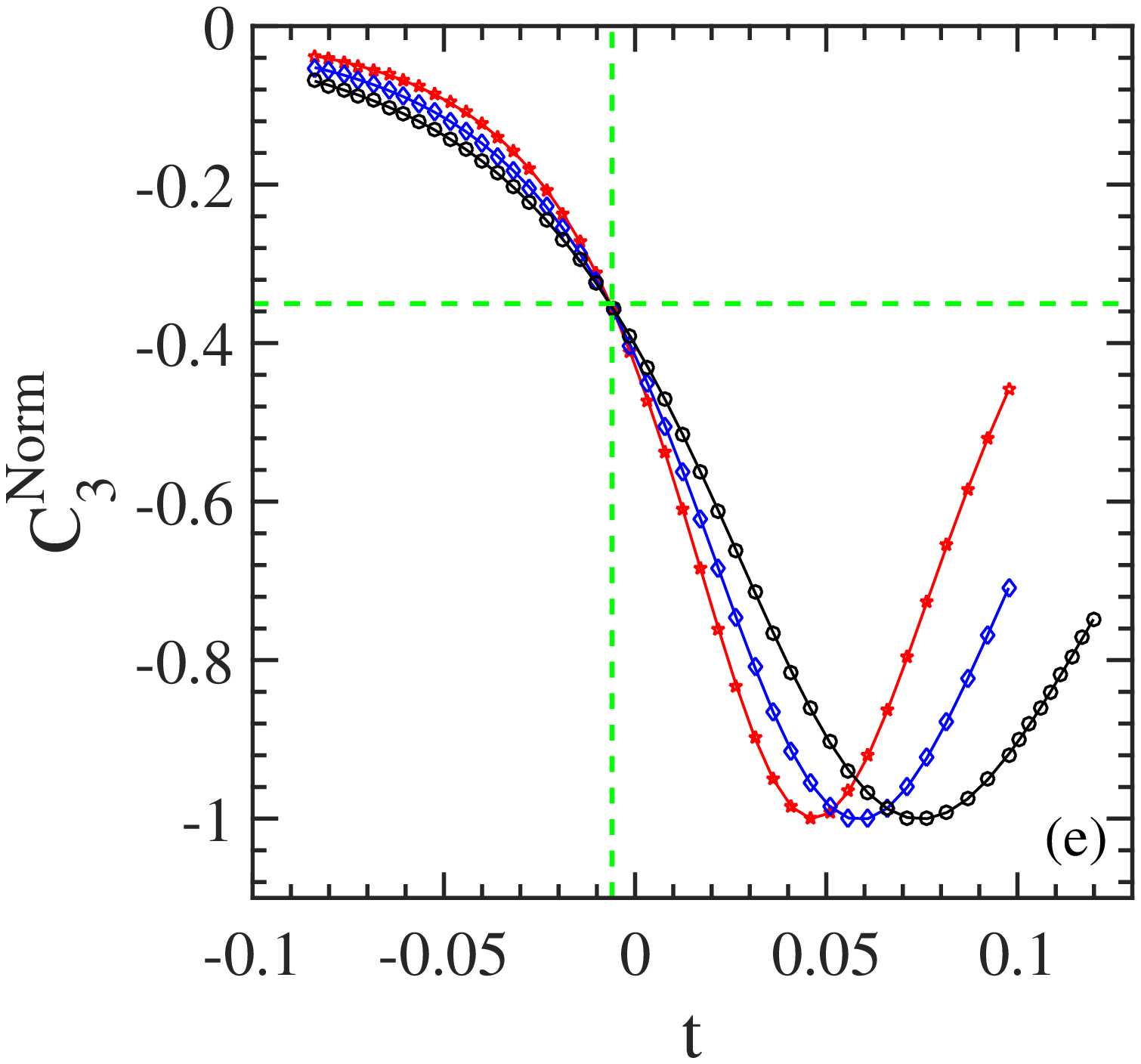}
    \includegraphics[width=0.3\textwidth]{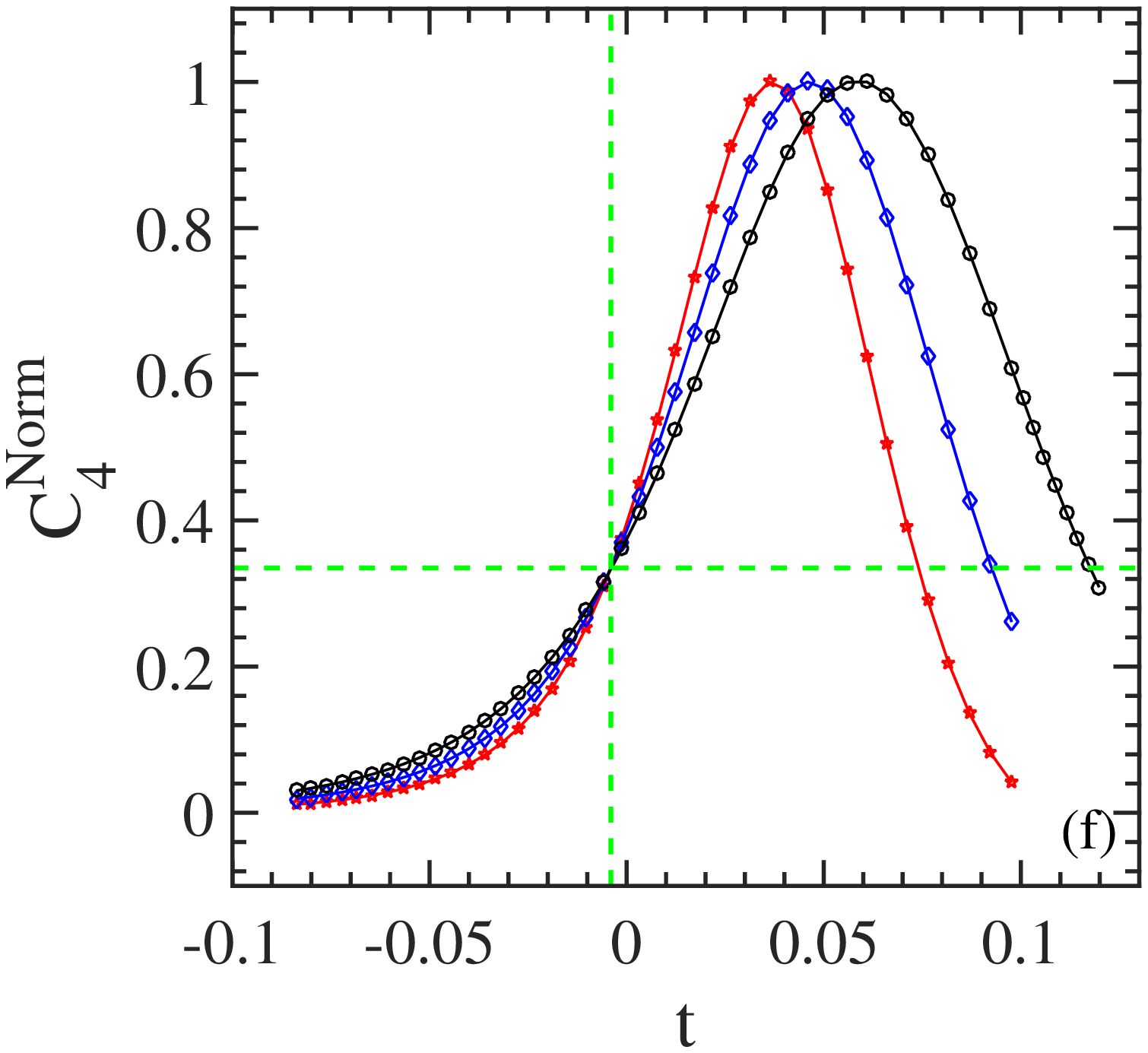}
    \caption{\label{Fig. 3}(Color online). Temperature dependence of $\kappa_2^{Norm}$ (a), $\kappa_3^{Norm}$ (b) and $\kappa_4^{Norm}$ (c), $C_2^{Norm}$ (d), $C_3^{Norm}$ (e) and $C_4^{Norm}$ (f) at three different values of external magnetic fields, $H = 0.05$, $0.07$ and $0.1$, in the three-dimensional Ising model simulated by Monte Carlo method. The cross point of green dashed lines is the fixed point for the upper panel. The positions of the green dashed lines in the lower panel are keep consistent with that in the upper panel.}
\end{figure*}

By Monte Carlo simulation method, the fixed point behavior is tested in finite-size systems at three different values of external magnetic fields. Because of the finite-size effects, the temperature dependence curves of the cumulants will shift to the higher temperature side until the system size is bigger enough and sufficiently converge to the thermodynamic limit. The typical size is determined by the saturation of size dependence of an observable at a given magnetic field~\cite{systemsize}. For cumulants up to the fourth-order at three different magnetic fields $H=0.05$, $0.07$, and $0.1$, lattice sizes $L=14$, $12$ and $10$ is sufficient to converge to the thermodynamic limit, respectively. For each value of $H$, the simulations are performed at $5$ values of inverse temperature $J/T= 0.202$, $0.212$, $0.222$, $0.232$ and $0.242$ near the critical temperature $T_c/J \approx 4.51$, where the value of interaction energy $J$ is set to $1$. The Wolff cluster algorithm is used with the helical boundary conditions~\cite{Wolff}. At each pair of $(H, J/T)$, $48$ million independent configurations are generated and used in a Ferrenberg-Swendsen reweighting analysis to calculate observables at intermediate temperature values~\cite{FS}.

Results of the second- to fourth- order normalized cumulants and factorial cumulants are shown in Fig.~3(a) to 3(f), respectively. It is clear that the fixed point behavior in temperature dependence of $\kappa_2^{Norm}$, $\kappa_3^{Norm}$ and $\kappa_4^{Norm}$ still exists as shown in Fig.~3(a) to Fig.~3(c). The cross point of the two green dashed line shows the position of the fixed point. The corresponding temperature is about one percent lower than the critical one. What is more, the higher the order of the cumulants, the closer of the fixed point to the critical temperature.  In addition, the values of the normalized cumulants at fixed points are different from those in the parametric representation. This can be caused by the finite-size system, the choice of function $h(\theta)$ and the different quantitative temperature dependence of $\kappa_n$ in Monte Carlo simulation and the parametric representation.

\begin{figure*}[htb]
\centering
    \includegraphics[width=0.35\textwidth]{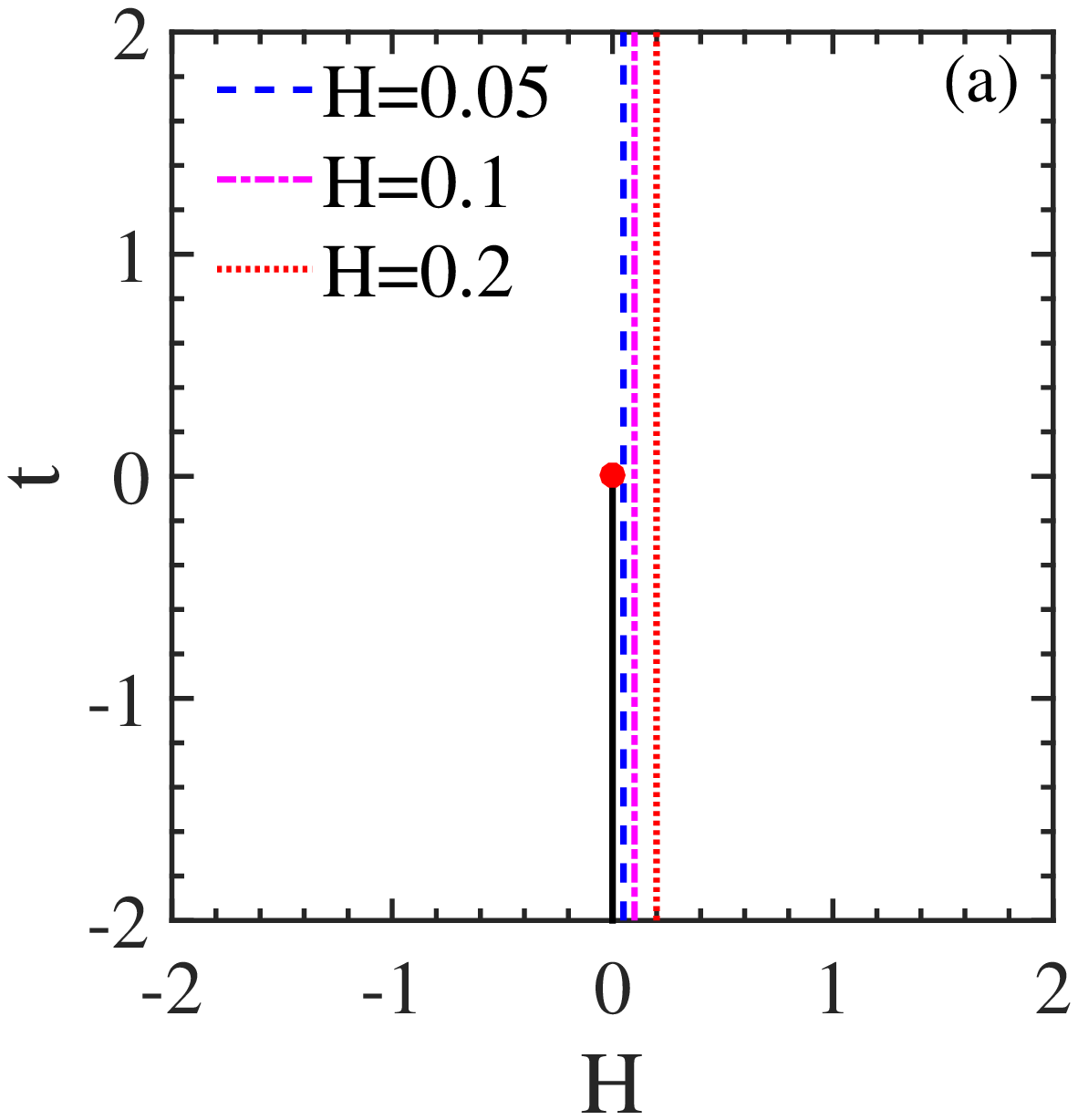}
    \includegraphics[width=0.4\textwidth]{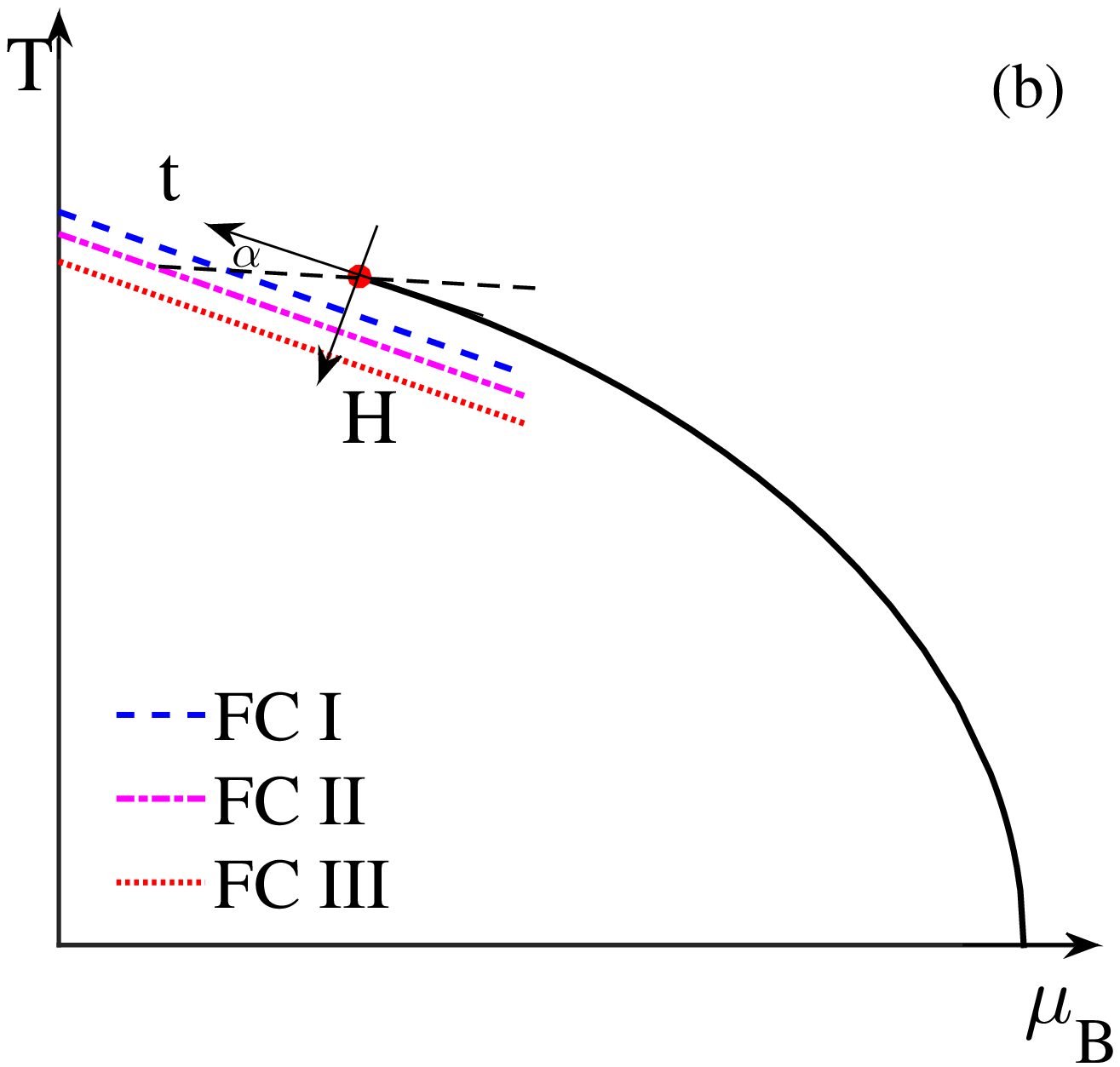}
    \caption{\label{Fig. 4}(Color online). Phase diagram on $t-H$ plane of the three-dimensional Ising model (a). Sketch of the $t-H$ axes mapped onto the QCD $T-\mu_B$ plane (b). The black solid line and the red point are the first-order phase transition line and the critical point, respectively. The $t$ axis is tangential to the QCD phase boundary at the critical point. The $H$ direction is set to be perpendicular to the $t$ axis. The three lines parallel to the phase boundary from left to right ($H = 0.05, 0.1$ and $H$ = 0.2) of Ising model (a) are mapped to the three freeze-out curves which is parallel to the $t$ axis from up to down (FC I, FC II and FC III) in QCD $T-\mu_B$ plane (b).}
\end{figure*}

Comparing temperature dependence of normalized factorial cumulants as showed in Fig.~3(d) to 3(f) with the same order normalized cumulants showed in Fig.~3(a) to Fig.~3(c), respectively, there are no significant difference between them. That is because the temperatures are close to the critical one. This result is consistent with that in Ref.~\cite{panx, Phys. Rev. C.93.034915} that these two kinds of cumulants can not be distinguished in the vicinity of the critical point. So it is not hard to understand that except $C_2^{Norm}$, there are both obvious fixed point behavior in the temperature dependence in $C_3^{Norm}$ and $C_4^{Norm}$ as showed in Fig.~3(e) and 3(f). The positions of the green dashed lines are set as the same with that in the same order normalized cumulants. It is not hard to infer that the temperature of fixed point in the normalized factorial cumulants is almost the same with that in the normalized cumulants.

All in all, at different external magnetic fields in finite-size systems of the three-dimensional Ising model, temperature dependence of the normalized cumulants or factorial cumulants (at least from the third-order) form a fixed point just about one percent distance to the critical temperature. Comparing with positions of the peak and sign change, the fixed point of normalized cumulants or factorial cumulants is much closer to the critical temperature.

\section{Fixed point behavior in the energy dependence of normalized cumulants and factorial cumulants}

\begin{figure*}[htb]

\centering
    \includegraphics[width=0.3\textwidth]{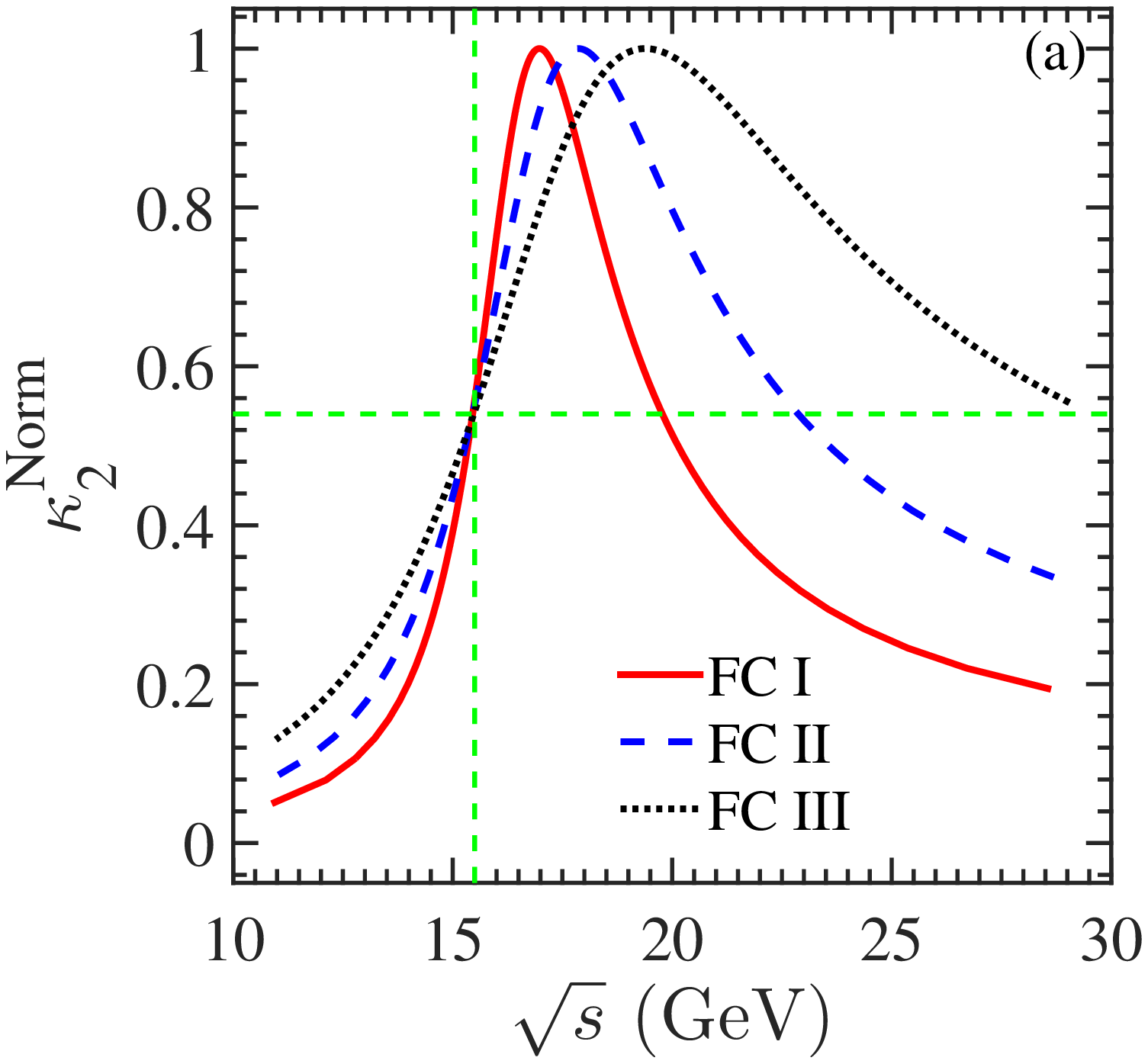}
    \includegraphics[width=0.3\textwidth]{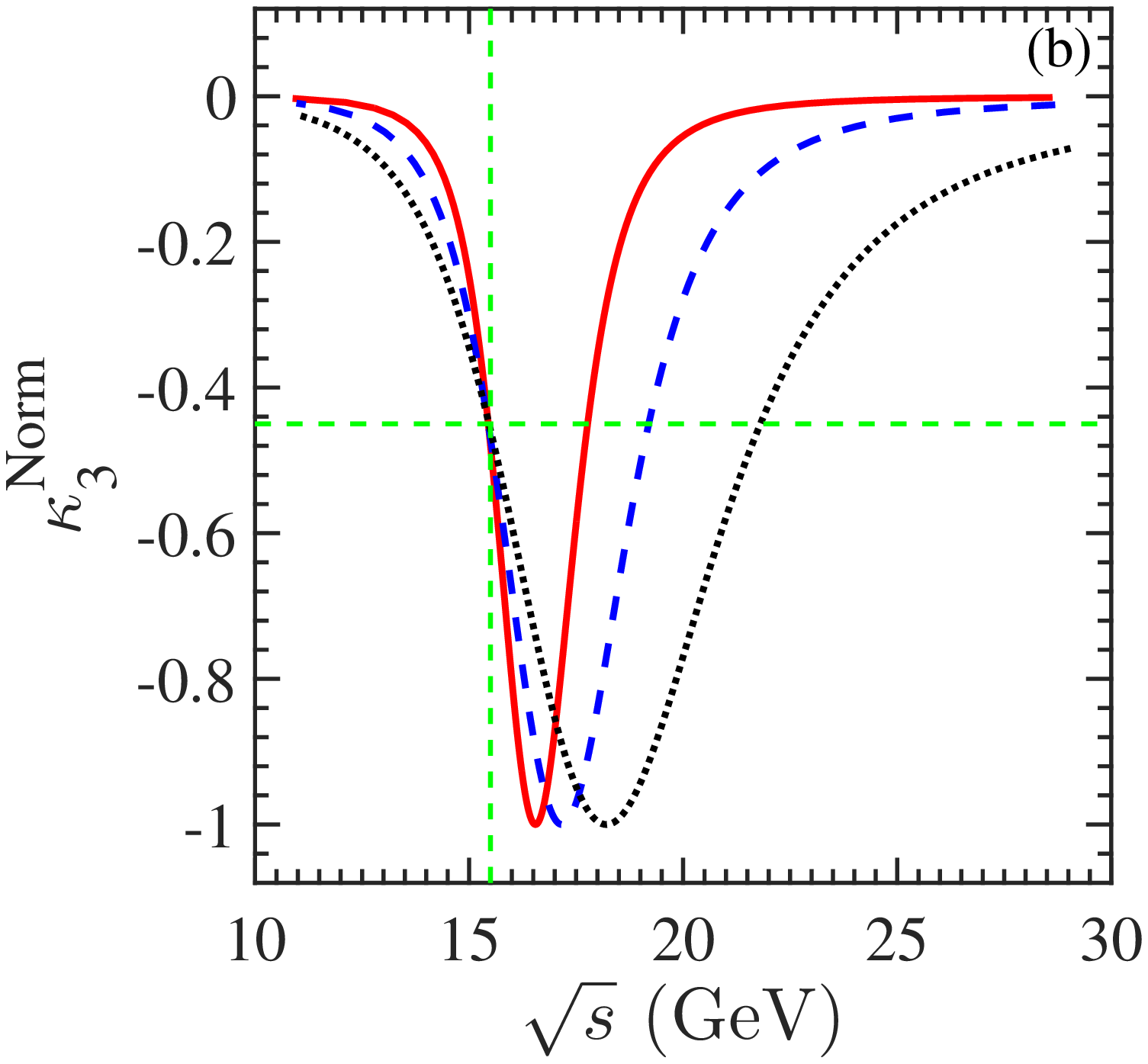}
    \includegraphics[width=0.3\textwidth]{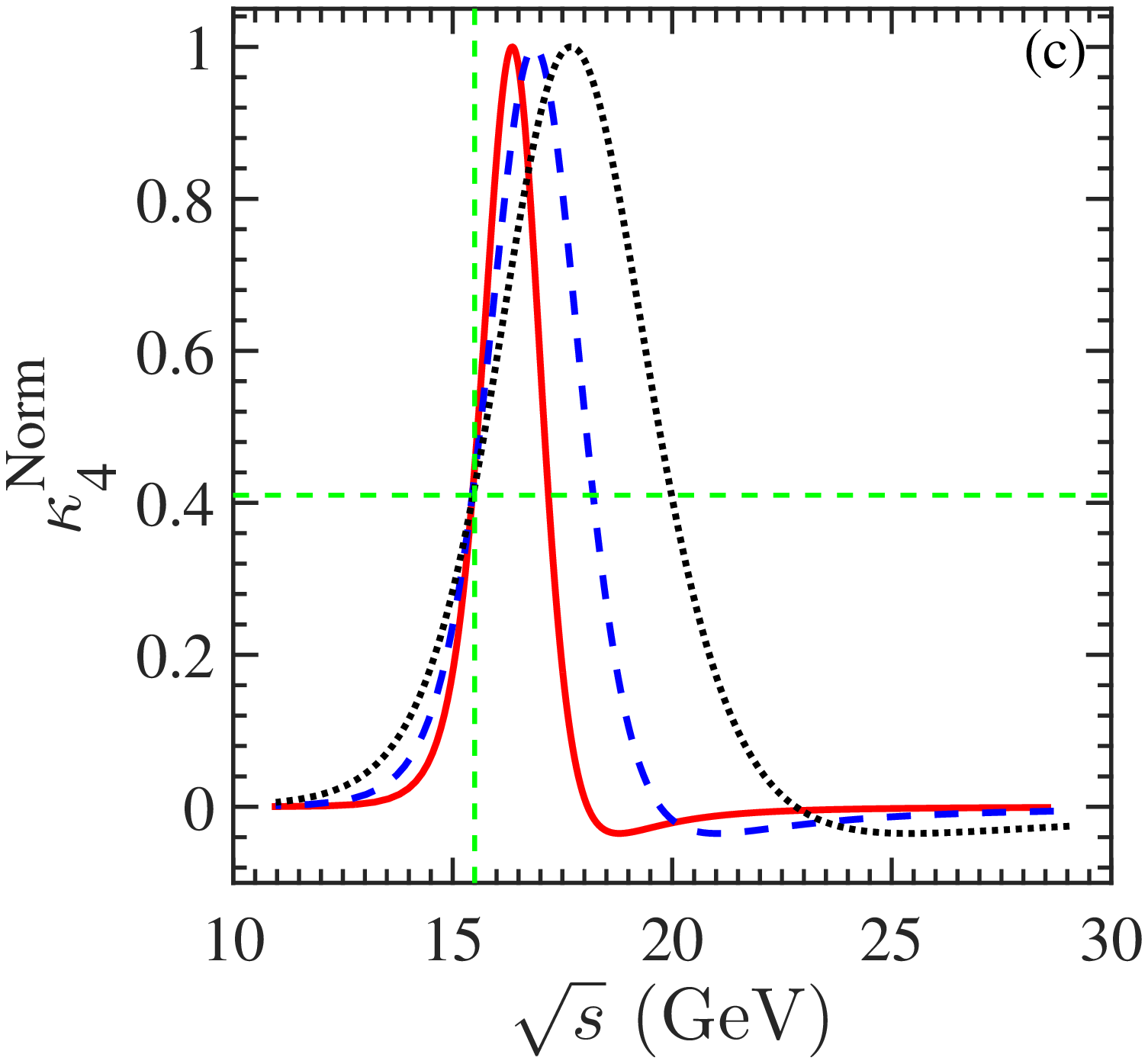}
     \includegraphics[width=0.3\textwidth]{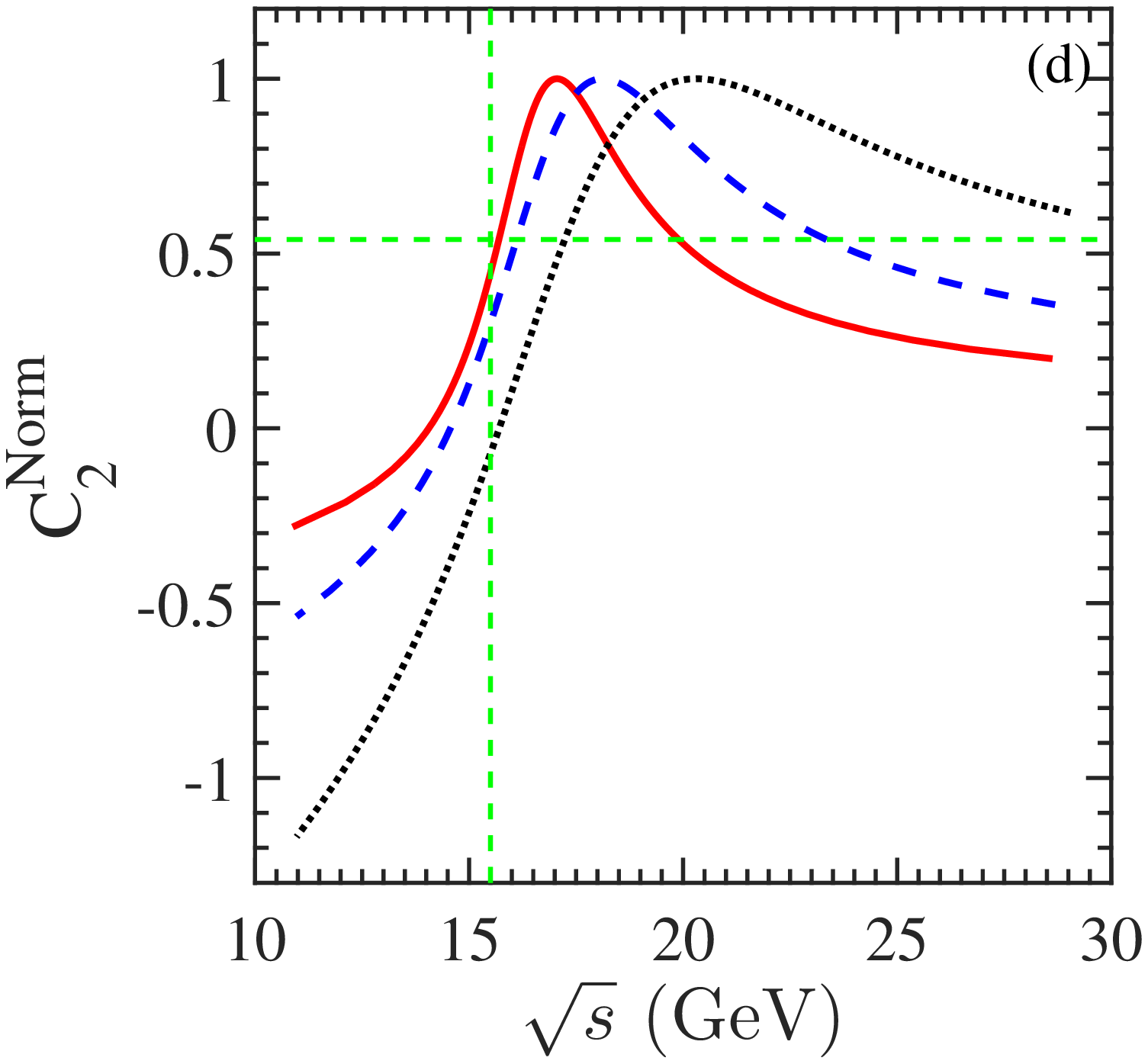}
    \includegraphics[width=0.3\textwidth]{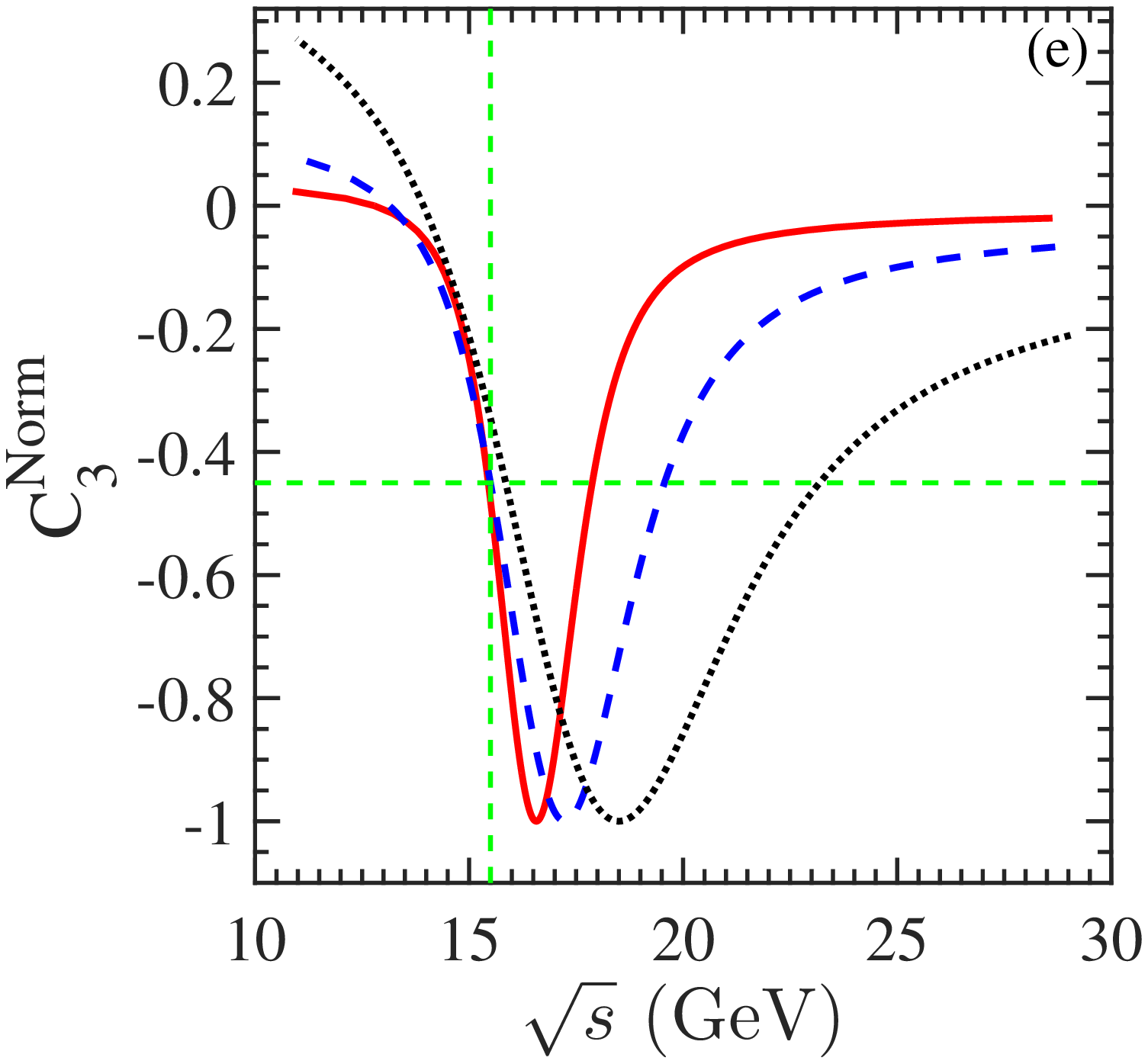}
    \includegraphics[width=0.3\textwidth]{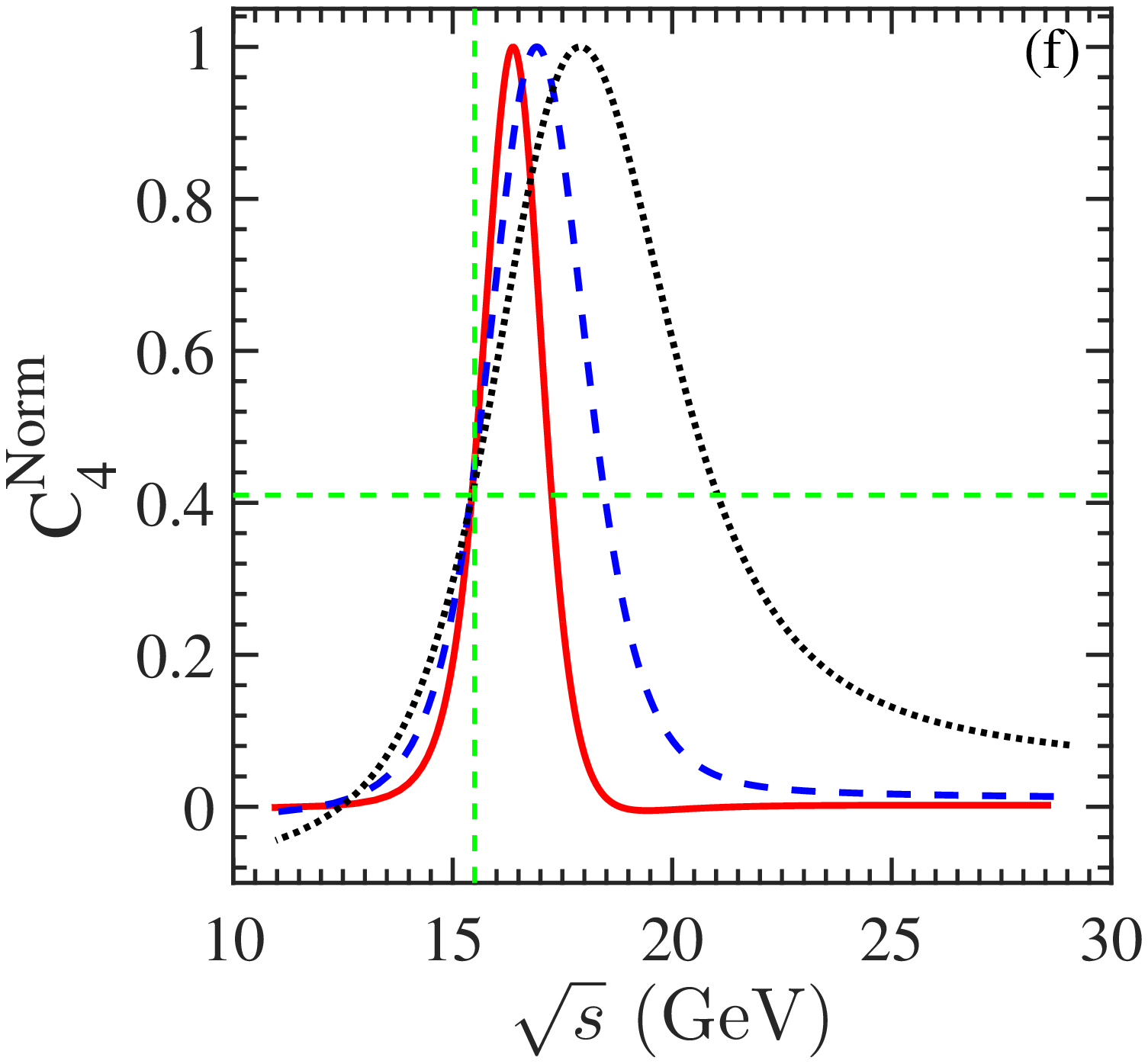}
    \caption{\label{Fig. 5}(Color online). Energy dependence of $\kappa_2^{Norm}$ (a), $\kappa_3^{Norm}$ (b) and $\kappa_4^{Norm}$ (c), $C_2^{Norm}$ (d), $C_3^{Norm}$ (e) and $C_4^{Norm}$ (f) along three different freeze-out curves as showed in Fig.~4(b). The vertical green dashed lines show the energy ${\sqrt s}_c$ corresponding to $\mu_{Bc}$. The horizontal green dashed lines show the value of the normalized cumulants at the fixed point for the upper panel. The positions of the green dashed lines in the lower panel are keep consistent with that in the upper panel.}
\end{figure*}

In the current study we focus on the equilibrium properties of the cumulants and factorial cumulants. Thus the non-equilibrium effects are not taken into account~\cite{Phys. Rev. C.92.034912}.

In order to apply the results of this paper to the heavy-ion collision experiments to search for the QCD critical point, it is essential to specify the map between the Ising variables $t$, $H$ to the QCD variables temperature $T$, baryon chemical potential $\mu_B$. The $t$ axis is tangential to
the first-order phase transition line at the QCD critical point. The angle between the horizontal (fixed $T$) lines on the QCD phase diagram and $t$ axis is $\alpha$. For simplicity, we assume that the $H$ axis is perpendicular to the $t$ axis after the map to the $T-\mu_B$ plane which has been studied in Ref.~\cite{linearmap3}. Then a linear mapping relations can be got as follows:
\begin{equation}\label{linear mapping}
\begin{split}
&\frac{T-T_{cep}}{\Delta T}=-\cos\alpha\frac{H}{\Delta H}+\sin\alpha \frac{t}{\Delta t},\\
&\frac{\mu_B-\mu_{Bc}}{\Delta \mu_B}=-\sin\alpha\frac{H}{\Delta H}-\cos\alpha \frac{t}{\Delta t}.\\
\end{split}
\end{equation}
Where $T_{cep}$, $\mu_{Bc}$ represent the temperature and baryon chemical potential at the QCD critical point. $\Delta T$ and $\Delta \mu_B$ denote the width of the critical regime in the QCD phase diagram. Because the location of the critical point and the width of the critical regime for QCD are not known, the suggestion that $\Delta \mu_B \approx 0.1$ GeV from model calculations~\cite{Phys. Rev. D.67.014028} and lattice QCD calculations~\cite{Phys. Rev. D.78.114503} is used. We set $\Delta \mu_B = 0.1$ GeV and $\mu_{Bc} = 0.25$ GeV as was done in Ref.~\cite{Phys. Rev. C.92.034912}.

$\Delta H$ and $\Delta t$ denote the width of the critical regime in the Ising variables. For simplicity, we set $\Delta H=0.4$ and $\Delta t=2$, respectively (the fixed point behavior is not sensitive to the width of the critical regime in the Ising variables. For more information to define the critical regime, to see Ref.~\cite{Phys. Rev. C.92.034912}).

At last, the freeze-out curve is assumed below the crossover/first-order phase transition line. An empirical parametrization of the heavy-ion-collision data from Ref.~\cite{Phys. Rev. C.73.034905} can be used to describe the freeze-out curves,
\begin{equation}\label{freeze-out curve}
T_f(\mu_B)=a-b{\mu_B}^2-c{\mu_B}^4.
\end{equation}
Where $a=0.166$ GeV, $b=0.139$ GeV$^{-1}$, $c=0.053$ GeV$^{-3}$. At a small range of $\mu_B$ (0.15 GeV < $\mu_B$ < 0.35 GeV), $T_f$ is approximately viewed varying linearly with $\mu_B$ in this study. The angle between the straight line of $T_f(\mu_B)$ and the horizontal (fixed $T$) line on the QCD $T-\mu_B$ plane is very small. For simplicity, we assume the freeze-out curve approximately parallel to the $t$ direction which has been mapped to the QCD phase diagram.

For straightforward, phase diagram of the three-dimensional Ising model on the $t-H$ plane and one possible sketch of the $t-H$ axes mapped onto the $T-\mu_B$ plane of QCD are showed in Fig.~4(a) and 4(b), respectively. Thus three lines parallel to the $t$ axis from left to right at three different values of $H$ in Fig.~4(a) can be simply mapped to three freeze-out curves from up to down in QCD as showed in Fig.~4(b).

Base on the mapping and using Eq.~\eqref{linear mapping}, the temperature dependence of normalized cumulants and factorial cumulants at the three different $H$ can be converted to the $\mu_B$ dependence of normalized cumulants and factorial cumulants along the three different freeze-out curves.

Turn to the heavy-ion collision experiments, using the energy ($\sqrt s$) dependence of $\mu_B$ given in Ref.~\cite{Phys. Rev. C.73.034905},
\begin{equation}\label{energy dependence of mu}
\mu_B(\sqrt s)=\frac{d_0}{d_1\sqrt s +1},
\end{equation}
where $d_0=1.308$ GeV, $d_1=0.273$ GeV$^{-1}$, one can get the energy dependence of normalized cumulants and factorial cumulants along the three different freeze-out curves.

Supposing the angel $\alpha=3^{\circ}$, based on the parametric representation of the Ising model, energy dependence of the second- to fourth-order normalized cumulants and factorial cumulants along the three freeze-out curves are showed in Fig.~5(a) to 5(f), respectively. The vertical green dashed lines show the critical energy ${\sqrt s}_c = 15.5$ GeV which is corresponding to $\mu_{Bc} = 0.25$ GeV at the QCD critical point through Eq.~\eqref{energy dependence of mu}. It is clear that, the fixed point behavior exists at ${\sqrt s}_c$ in the energy dependence of $\kappa_2^{Norm}$ to $\kappa_4^{Norm}$ as showed in Fig.~5(a) to 5(c), respectively. The values of the normalized cumulants at the fixed point showed by the horizontal green dashed line are a little changed compared to the values given by Eq.~\eqref{normalized cumulants at t=0} because of the mapping from the Ising variables to the QCD variables.

For the normalized factorial cumulants showed in Fig.~5(d) to 5(f), the fixed point behavior occurs from the fourth-order one, and its position is consistent with that in $\kappa_4^{Norm}$.

The fixed point behavior in the energy dependence of the normalized cumulants is derived directly from the linear mapping in Eq.~\eqref{linear mapping}, where $\Delta H$, $\Delta t$ and the angle $\alpha$ are all set a fixed value in this paper. The influence of these three parameters on the fixed point behavior should be explained. The fixed point behavior still exists as the variation of these three parameters. Different values of $\Delta H$ and $\Delta t$ almost do not change the energy at the fixed point. They just influence the range of the energy (the range of $\mu_B$) after the mapping. Small values of $\alpha$ (like $3^{\circ}$ used in this paper) has little influence on the fixed point behavior. While soaring values of $\alpha$ not only change the range of the energy, but also shift the fixed point away from ${\sqrt s}_c$ (but one should notice that a small value for $\alpha$ should be more closer to the truth here).

One other problem should be discussed is that how can one get different freeze-out curves in the heavy-ion collisions? In fact, the centrality dependence of the chemical freeze-out temperature and baryon chemical potential has been studied in Ref.~\cite{Phys. Rev. C.71.054901, Advances in High Energy Physics.2021.6611394}. Although the chemical freeze-out temperature does not vary much with centralities, the temperature interval between the three different freeze-out curves can be very small. If we set $T_{cep} = 1.8$ GeV and $\Delta T = T_{cep}/8$ which has been used in Ref.~\cite{Phys. Rev. C.92.034912}, the critical regime of QCD temperature is $\Delta T$ = 0.0225 GeV. When the external magnetic field $H$ changes from 0.05 to 0.2 at the same $t$, after mapping to the QCD variables through Eq.~\eqref{linear mapping}, the freeze-out temperature interval is just about $0.0084$ GeV. What is more, baryon chemical potential increases from peripheral to the most central collisions~\cite{Advances in High Energy Physics.2021.6611394}. It is enough for one to get different freeze-out curves at different centralities. So the centrality controlling the freeze-out curves in the QCD phase diagram may play a similar role of the external magnetic field $H$ of the Ising model.

Under the mapping from the three-dimensional Ising model to QCD, the fixed point behavior may be expected in the energy dependence of normalized net-proton (factorial) cumulants in heavy-ion collision experiments. This feature can be used to locate the QCD critical point.

\section{Summary}

By using the parametric representation of the three-dimensional Ising model, temperature dependence of the second- to fourth-order cumulants and factorial cumulants of the order parameter is studied. The qualitative behavior of temperature dependence of cumulants does not change with the varying external magnetic field in the vicinity of the critical temperature. So does that of the factorial cumulants.

The fixed point behavior in temperature dependence of normalized cumulants at the critical temperature for different magnetic fields is deduced and showed.

By Monte Carlo simulation of the three-dimensional Ising model, the fixed point behavior in the temperature dependence of normalized second- to fourth-order cumulants is checked in finite-size systems. The fixed point behavior still exists just about one percent distance to the critical temperature, which is much closer to the critical temperature than the peak structure or sign change showed in the temperature dependence of the cumulants.

For the normalized factorial cumulants, the fixed point behavior is also survived at least from the fourth order cumulants both in the parametric representation and finite-size systems which reflect the fact that the critical behavior of factorial cumulants is dominant by the corresponding cumulants. The higher the order of the factorial cumulant, the more dominant role of the same order cumulant in its critical behavior.

Through a mapping from the three-dimensional Ising model to QCD, the fixed point behavior is also found in the energy dependence of the normalized cumulants (or fourth-order factorial cumulants) along different freeze-out curves. The fixed point is very close to the critical energy (corresponding to the baryon-chemical potential at the QCD critical point). It should be promising for the method to be applicable to locate the QCD critical point in heavy-ion collision experiments.

More generally it must be emphasized that all of the results here rely on the equilibrium of the system. Whether the fixed point behavior survived in the non-equilibrium cumulants needs further studies. What is more, further studies of different ways of mapping from the Ising model to QCD will be helpful.

\vskip 0.5cm
The author acknowledges fruitful discussions with Yuanfang Wu, MingMeiXu and Lizhu Chen

\ed